%% file: vortextunnelling-v4.tex
\documentclass[pra,preprint,showpacs]{revtex4-1}

\usepackage{amsmath,latexsym,amssymb,amsfonts}
\usepackage[usenames,dvipsnames]{color}
\usepackage[colorlinks]{hyperref}

\usepackage{graphicx}
\usepackage[applemac]{inputenc}
\usepackage[T1]{fontenc} 

\newcommand{\beq}{\begin{equation}}
\newcommand{\eeq}{\end{equation}}
\newcommand{\bea}{\begin{eqnarray}}
\newcommand{\eea}{\end{eqnarray}}

\newcommand{\eps}{\epsilon}
\newcommand{\Ga}{\Gamma}
\newcommand{\hR}{{\hat R}}
\newcommand{\hE}{{\hat E}}
\newcommand\ltwid{\mathrel{
 \raise.3ex\hbox{$<$\kern-.75em\lower1ex\hbox{$\sim$}}}}

\begin{document}

\preprint{UdeM-GPP-TH-12-207}
\preprint{CQUeST-2013-0617}
\preprint{YITP-13-68}

\title{Tunneling decay of  false vortices}
\author{Bum-Hoon Lee$^{a,b}$}
\email{bhl@sogang.ac.kr}
\author{Wonwoo Lee$^{b}$}
\email{warrior@sogang.ac.kr}
\author{Richard MacKenzie$^{c}$}
\email{richard.mackenzie@umontreal.ca}
\author{M.~B.~Paranjape$^{c}$}
\email{paranj@lps.umontreal.ca}
\author{U.~A.~Yajnik$^{d}$}
\email{yajnik@iitb.ac.in}
\author{Dong-han Yeom$^{b,e}$}
\email{innocent.yeom@gmail.com}
\affiliation{$^a$Department of Physics and BK21 Division, Sogang University, Seoul 121-742, Korea}
\affiliation{$^b$Center for Quantum Spacetime, Sogang University, Seoul 121-742, Korea}
\affiliation{$^c$Groupe de Physique des Particules, Universit\'{e} de Montr\'{e}al, C.~P.~6128, Succursale Centre-ville, Montreal, Qu\'{e}bec, Canada, H3C 3J7}
\affiliation{$^d$Department of Physics, Indian Institute of Technology Bombay, Mumbai, India}
\affiliation{$^e$Yukawa Institute for Theoretical Physics, Kyoto University, Kyoto 606-8502, Japan}

\begin{abstract}
We consider the decay of vortices trapped in the false vacuum of a theory of scalar electrodynamics in 2+1 dimensions.    The potential is inspired by models with intermediate symmetry breaking to a metastable vacuum that completely breaks a $U(1)$ symmetry, while in the true vacuum the symmetry is unbroken.  The false vacuum is unstable through the formation of true vacuum bubbles; however, the rate of decay can be extremely long.  On the other hand,  the false vacuum can contain metastable vortex solutions.  These vortices contain the true vacuum inside in addition to a unit of magnetic flux and the appropriate topologically nontrivial false vacuum outside.  We numerically establish the existence of vortex solutions which are classically stable; however, they can decay via tunneling.  In general terms, they tunnel to a configuration which is a large, thin-walled vortex configuration that is now classically unstable to the expansion of its radius.  We compute an estimate for the tunneling amplitude in the semi-classical approximation. We believe our analysis would be relevant to superconducting thin films or superfluids.
\end{abstract}

\maketitle

\newpage

\section{Introduction \label{sec-1}}

Vortices are topological solitons in a spontaneously broken $U(1)$ gauge theory of a complex scalar field $\phi$ in two space dimensions. The potential is minimized for a nonzero value of $|\phi|$, so the space of vacua is a circle, as is spatial infinity. In a finite-energy field configuration, $\phi$ must tend towards a vacuum at infinity, but its phase can change by $2\pi$ as the polar angle changes by $2\pi$, resulting in a vortex. (More generally, the phase can change by $2\pi n$, resulting in an $n$-vortex, the integer $n$ being the winding number of the configuration.) Continuity of $\phi$ dictates that it must vanish somewhere (normally taken to be the origin); thus, the core of the vortex has nonzero energy density. Finiteness of energy also requires that the vortex have a (quantized) magnetic flux in its core.

In such a model in three space dimensions, the soliton becomes a one-dimensional topological defect, as described by Abrikosov \cite{abr57} and by Nielsen and Olesen  \cite{no00}. These objects exist in many realistic models in particle physics; in the cosmology of such models, they are formed during phase transitions in the early universe \cite{kibble} and are known as cosmic strings.  In condensed matter physics, they appear as vortex lines in type-II superconductors \cite{gl}.  

Vortices and strings also appear in global (that is, non-gauged) models with a complex scalar field alone. In a condensed matter context, global strings correspond to vortex lines in superfluid $^4$He, where the scalar field is the condensate wavefunction \cite{leggett}.  A global vortex has a logarithmically divergent energy; hence in the two dimensional context they must come in vortex-antivortex pairs.  In three dimensions the global vortex string must be finite and joined to form a closed loop.  This last condition can be relaxed if gravitation is also taken into account and the condition becomes simply that the energy be finite within a Hubble radius \cite{akvs}.

In this paper we will restrict ourselves to the case of 2+1-dimensional gauge theory vortices.  Due to the form of the potential (see (\ref{potential}) below), there is an important difference between the vortices in our model and those discussed in \cite{abr57,no00}.  In both cases the field interpolates between zero at the origin and a nonzero value at spatial infinity, but in standard vortices this corresponds to going from a maximum of the potential to the true vacuum at infinity.  Our vortices have the opposite behavior, in a sense, in that the scalar field goes from the true vacuum at the origin to a symmetry-breaking false vacuum at infinity.

The scalar field profile as a function of $r$ is determined by the equations of motion, of course, but two possible profiles can be identified. The first is a ``thin-wall'' profile, for which the scalar field remains at its central value $\phi=0$ inside the core of the vortex, then quickly jumps up to a value very close to the final symmetry-breaking value, which it then approaches exponentially. The second is a ``thick-wall'' or normal vortex, for which the scalar field varies throughout the core of the vortex. Thin-wall vortices are desirable as they allow certain simplifications in the analysis; however, they are harder to produce (that is, they arise in a smaller region of parameter space), as we will see.

It is not obvious that the potential we consider will give rise to classically stable vortices (whether thin-wall or thick-wall).  Indeed, simply expanding the interior region where $|\phi|\simeq0$ should eventually give rise to a lower energy.  It is possible, however, that there is an energy barrier separating the initial vortex configuration and any lower-energy configuration, giving rise to a classically stable vortex.  

\begin{figure}[ht]
\begin{center}
\includegraphics[width=3.in]{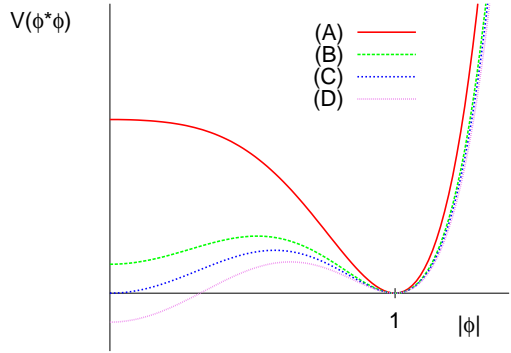}
\end{center}
\caption{(Color online) The potential (\ref{potential1}) varying from the standard symmetry-breaking form (A) to one with a symmetric true vacuum and symmetry-breaking false vacuum (D).
}
\label{fig1}
\end{figure}

In Figure \ref{fig1} we have drawn four possible examples of the scalar potential, starting with (A) which is the usual quartic symmetry-breaking potential with a maximum at $\phi=0$ and with vacuum at $|\phi |=1$ after a suitable rescaling, (B) where a metastable local minimum (false vacuum) is formed at $\phi=0$, (C) where this minimum becomes degenerate with the symmetry-breaking minimum, and (D) where the roles of the two vacua are reversed, with $\phi=0$ and $|\phi |=1$ becoming the true and false vacua, respectively.

In cases (A) and (B), it is easily shown that a vortex solution exists and is stable both classically and quantum mechanically.  To see this, note that the energy of the flux trapped in a vortex of core size $R$ makes a contribution of order $(\pi R^2/2)|\vec B|^2=\Phi^2/(\pi R^2)$, where $\Phi$ is the total flux, stabilizing the configuration against collapse.  On the other hand, the potential term will diverge as $\sim \pi R^2 V(0)$ (where in these cases $V(0)>0$) for large $R$, stabilizing the configuration against expansion.

The covariant derivative of the scalar field also contributes an energy linear in $R$, whether the configuration is a thin-wall or thick-wall vortex.  In the thin-wall case, for a large vortex, the dynamics governing the edge of the vortex should be independent of its size, and since the scalar field must vary from the value zero inside  to one outside, the linear distance over which it will do so should be independent of $R$.  Hence the gradient behaves as $\sim 1/\Delta$ for some constant $\Delta$ representative of the wall thickness.  Its square gives the energy density which is distributed  over a circumference of length  $\sim R$, giving the stated behavior.   For a thick-wall vortex, the gradient energy is distributed evenly over the size of the vortex, then it will behave as $\sim 1/R$; squaring and multiplying by the area of the core gives an $R$-independent contribution to the energy.  However, in this case, the potential is nonzero throughout the core, again giving a contribution $\sim  R^2$.  Thus in either of these cases, the energy in the scalar field stabilizes the configuration against infinite dilation.

Case (C)  is the critical case, but here also the vortex solution exists and is classically stable.  The potential energy from the regions where the scalar field is nonzero contributes as in cases (A) and (B), preventing infinite dilation, and as before  the magnetic flux prevents collapse of the vortex.  This vortex could tunnel quantum mechanically to an infinitely large and infinitely diluted vortex. However, the amplitude for such a transition probably vanishes.     Thus case (C) is not the boundary of classical stability and continuity suggests that for case (D) there will also exist classically stable vortex configurations as long as $V(0)$ is close enough to zero, and indeed we will see that this is the case.   This paper is concerned with the quantum mechanical disintegration of such vortices.  
 
This could potentially be important for the following reason. In a model whose potential is given by case (D), if the universe is trapped in the false vacuum with $|\phi | = 1$, then the vaccum will ultimately decay. Standard vacuum decay \cite{Coleman} is exponentially suppressed, so the universe could be trapped in a false vacuum for a very long time. But generically vortices will be formed in a symmetry-breaking phase transition. If this phase transition is followed by a second phase transition which restores the symmetry (for instance, going from case (A) to case (D)), then these vortices have a core which is true vacuum. Intuitively, this might speed up vacuum decay, since the vortex core provides a region of space where the scalar field already is where it ultimately will be; we might suspect therefore that vortices have an important effect on vacuum decay. Although this is not necessarily the case, as we will show below it is possible for vortex-mediated vacuum decay to be much faster than normal vacuum decay.

In the next section we present the model in detail and discuss vortex solutions in it. The two types of vortices mentioned above will be examined. In the potential we have chosen, the simplest with the desired vacuum structure, thin-wall vortices require large winding number. It is possible that this conclusion depends on the detailed form of the potential.

In the following section we discuss the decay of vortices via tunneling. The analog of the bounce (instanton) solution of conventional vacuum decay will be discussed, and expressions for its action will be given for three cases (thick-wall vortices, thin-wall vortices and vortices in the so-called dissociation limit, a point in parameter space where thin-wall vortices become unstable). Not surprisingly, in the latter case the presence of vortices will indeed have an important effect on the decay of a false vacuum.

\section{False-vacuum vortex solution \label{sec-2}}

\subsection{Set-up}\label{subsec-setup}

We consider the abelian Higgs model (spontaneously broken scalar electrodynamics) with a modified scalar potential.  The Lagrangian density of the model has the form
\beq
{\cal L} = - \frac{1}{4} F_{\mu\nu}F^{\mu\nu} + (D_{\mu}\phi)^*(D^{\mu}\phi)-V(\phi^*\phi),
\label{lagran01}
\eeq
where
$F_{\mu\nu} = \partial_{\mu}A_{\nu} - \partial_{\nu}A_{\mu}$ and
$D_{\mu}\phi = (\partial_{\mu} - ie A_{\mu})\phi$.
The potential is a sixth-order polynomial in $\phi$
\cite{kpy, pjs}, written
\beq
V(\phi^*\phi) = \lambda(|\phi |^2-\eps v^2) (|\phi |^2-v^2)^2. \label{potential1}
\eeq
Note that the Lagrangian is renormalizable in 2+1 dimensions. The fields $\phi$ and $A_\mu$, the vacuum expectation value $v$ and the charge $e$ all have mass dimension 1/2, while the constants $\lambda$ and $\eps$ are dimensionless parameters controlling the
strength of the self-interaction of the scalar field.  The value of $\eps$ determines the shape of the potential (see Fig.~\ref{fig1}), the case of interest (D) corresponding to $0<\eps<1$ (beyond which there is no longer a barrier between $|\phi |=0$ and $|\phi | =1$). The potential energy density of the false vacuum $|\phi |=v$ vanishes, while that of the true vacuum is $V(0)=-\lambda\eps v^6$.   Instanton-type solutions corresponding to true vacuum bubbles (in a sea of the false vacuum) will then have finite (Euclidean) action.

\begin{figure}
\begin{center}
\includegraphics[width=3.in]{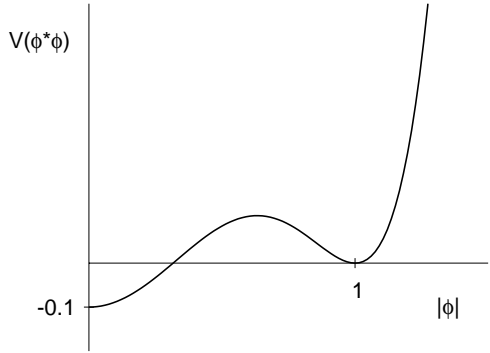}
\end{center}
\caption{The rescaled potential (\ref{potential}) with $\eps=0.1$.
} \label{fig:fig01}
\end{figure}

After rescaling by appropriate powers of $v$ and $\lambda$ so that all fields, constants and the spacetime coordinates are dimensionless, the Lagrangian density is still given by (\ref{lagran01}), multiplied by an overall factor of $\lambda^{-1/2}$,  where now the potential is 
\beq
V(\phi^*\phi) = (|\phi |^2-\eps) (|\phi |^2-1)^2. \label{potential}
\eeq
The overall factor does not affect the equations of motion and for the quantum theory is absorbed into an appropriate redefinition of $\hbar$.  The potential for a value of $\eps$ in the range of interest is exhibited in Fig.~\ref{fig:fig01}.

As mentioned in the introduction, in a false-vacuum universe topologically nontrivial field configurations (vortices) exist. These configurations may or may not be classically stable, but even if classically stable they can tunnel quantum mechanically to configuration of the same energy with a large core of true vacuum which will then expand rapidly.

We will look for rotationally-symmetric solutions for $\phi$ and $A_{\mu}$ in
polar coordinates $(r$, $\theta$, $t)$. We use the following time-dependent ansatz
for a vortex of winding number $n$:
\beq
\phi(r, \theta, t) = f(r, t) e^{in\theta}, \qquad  A_{i}(r,
\theta, t)=-\frac{n}{e} \frac{\varepsilon^{ij}{r}_j }{r^2}a(r, t),
\label{ansatz}
\eeq
where $\varepsilon^{ij}$ is the
two-dimensional Levi-Civita symbol.

The energy functional for the vortex has the form
\beq
E[A_{\mu}, \phi]= \int d^2x \left[+\frac{1}{2}F_{0i}F_{0i} + \frac{1}{4}F_{ij}F_{ij} + (D_{0}\phi)^*(D_{0}\phi) + (D_{i}\phi)^*(D_{i}\phi) + V(\phi^*\phi) \right] .  \label{efun}
\eeq
Substituting (\ref{potential},\ref{ansatz}) into (\ref{efun}), we obtain
\beq
E = 2\pi \int^{\infty}_{0} dr\, r \left[ \frac{n^2 \dot{a}^2}{2 e^2r^2}
 + \frac{n^2a'^2}{2 e^2r^2} + \dot{f}^2 + f'^2 + \frac{n^2}{r^2}(1-a)^2f^2
+(f^2-\eps) (f^2-1)^2 \right] , \label{energyvortex}
\eeq
where the prime and dot denote differentiation with respect to $r$ and $t$, respectively. We see that there are three parameters at play: the electric charge $e$, the parameter $\epsilon$, and the winding number $n$.

\subsection{Static solution}\label{subsec-staticsolution}

The static vortex solution is the minimum of this functional (without the time-derivative terms); the variational field equations are
\bea
f'' + \frac{f'}{r} - \frac{n^2}{r^2} (1-a)^2 f 
-(f^2-1)(3f^2-(1+2\eps))f &=& 0, \label{asymptotic1}\\
a'' - \frac{a'}{r}+ 2 e^2(1-a)f^2 &=& 0. \label{asymptotic2}
\eea
The form of the functions $f(r)$ and $a(r)$ can be found numerically using the following boundary conditions:
\bea
f(r)\rightarrow 0,  \quad a(r) \rightarrow 0 \quad &{\rm as}& \quad r \rightarrow 0,
\label{boundary1} \\
f(r)\rightarrow 1,  \quad a(r) \rightarrow 1 \quad &{\rm as}& \quad r \rightarrow \infty .  \label{boundary2}
\eea
Conditions (\ref{boundary1}) are imposed for smoothness of the fields at $r=0$ while (\ref{boundary2}) are required for finiteness of the energy. More precisely, the behavior for small $r$ can be found by linearizing the equations, which indicates that $f\sim r^n$ and $a\sim r^2$ as $r\to0$. As $r\to\infty$, we write $f(r)=1-\xi(r)$ and $a(r)=1-\psi(r)$ and linearize in $\xi$ and $\psi$. These functions obey modified Bessel equations, and we find $\xi(r)\sim r^{-1/2}e^{-2\sqrt{1-\eps}\,r} $ and $\psi(r)\to r^{1/2}e^{-\sqrt{2}\,er}$ as $r\to\infty$ \cite{rubakov}.

\begin{figure}
\begin{center}
\includegraphics[width=7cm]{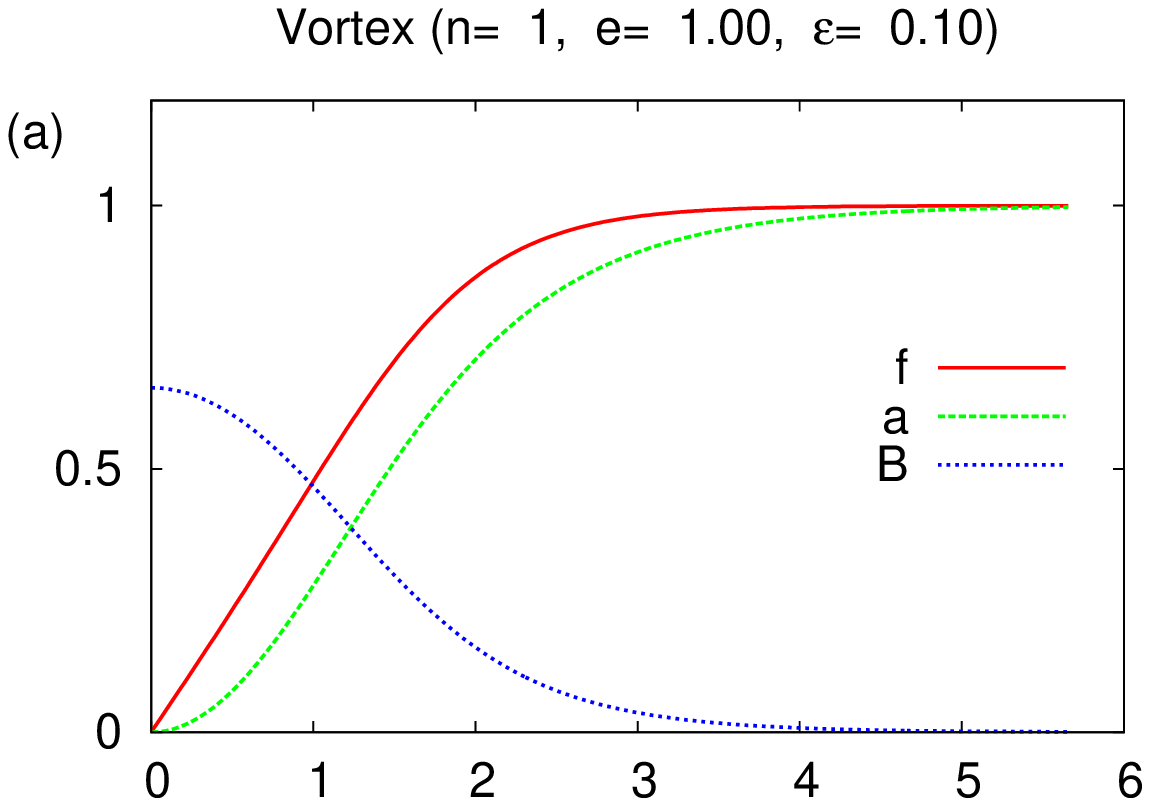}
\qquad
\includegraphics[width=7cm]{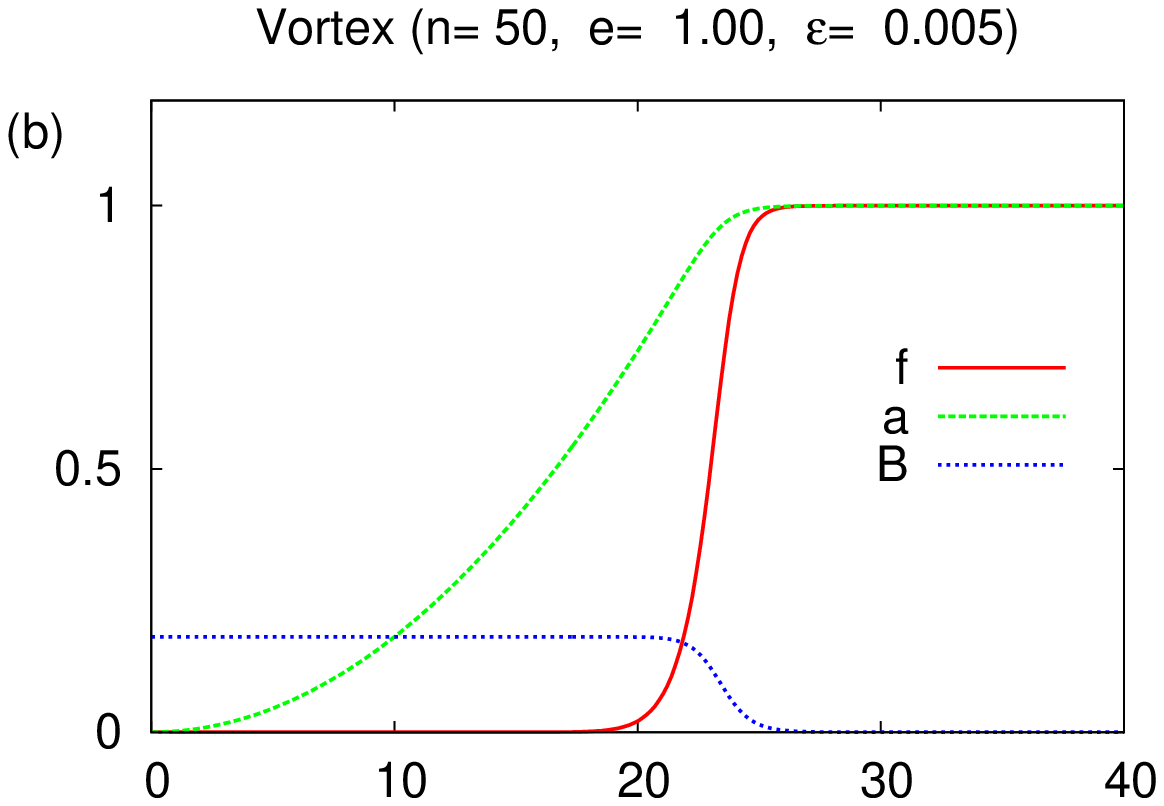}
\end{center}
\caption{(Color online) Vortex profile for (a) thick-wall and (b) thin-wall vortices. Displayed are the functions $f(r)$ and $a(r)$ and the magnetic field $B(r)=na'(r)/er$.} 
\label{fig:fig02}
\end{figure}

Numerical solutions for $f(r)$ and $a(r)$ are displayed in Fig.~\ref{fig:fig02}a for $n=e=1,\ \eps=0.1$. The vortex solution is classically stable.
 The asymptotic behavior (as $r \rightarrow 0$ and $r \rightarrow \infty$)
 of the profile functions
$f(r)$ and $a(r)$ is as expected. The solution has a thick-wall profile, unlike the case of usual vacuum bubbles studied in \cite{Coleman} which have a thin wall in the limit that the vacuum degeneracy splitting, controlled by the value of $\eps$, is very small.  But it should be noted that here we are looking for classically stable soliton solutions, and not instanton-type solutions analogous to the thin-wall vacuum bubbles found in \cite{Coleman}. Nonetheless, it will prove useful to have a thin-wall vortex solution since then tunneling can be analysed without recourse to numerical simulation, as indeed was the case in \cite{Coleman}. We have not found any such solutions with $n=1$. This may be a consequence of the form of the potential chosen, although we have yet to consider other potentials to see if thin-wall vortices with $n=1$ can be produced. However, with the potential (\ref{potential}),  thin-wall solutions do exist for sufficiently small $\eps$ and sufficiently large $n$ \cite{bol1,bol2,bol3}; one such solution is displayed in Fig.~\ref{fig:fig02}b.

\begin{figure}
\begin{center}
\includegraphics[width=7cm]{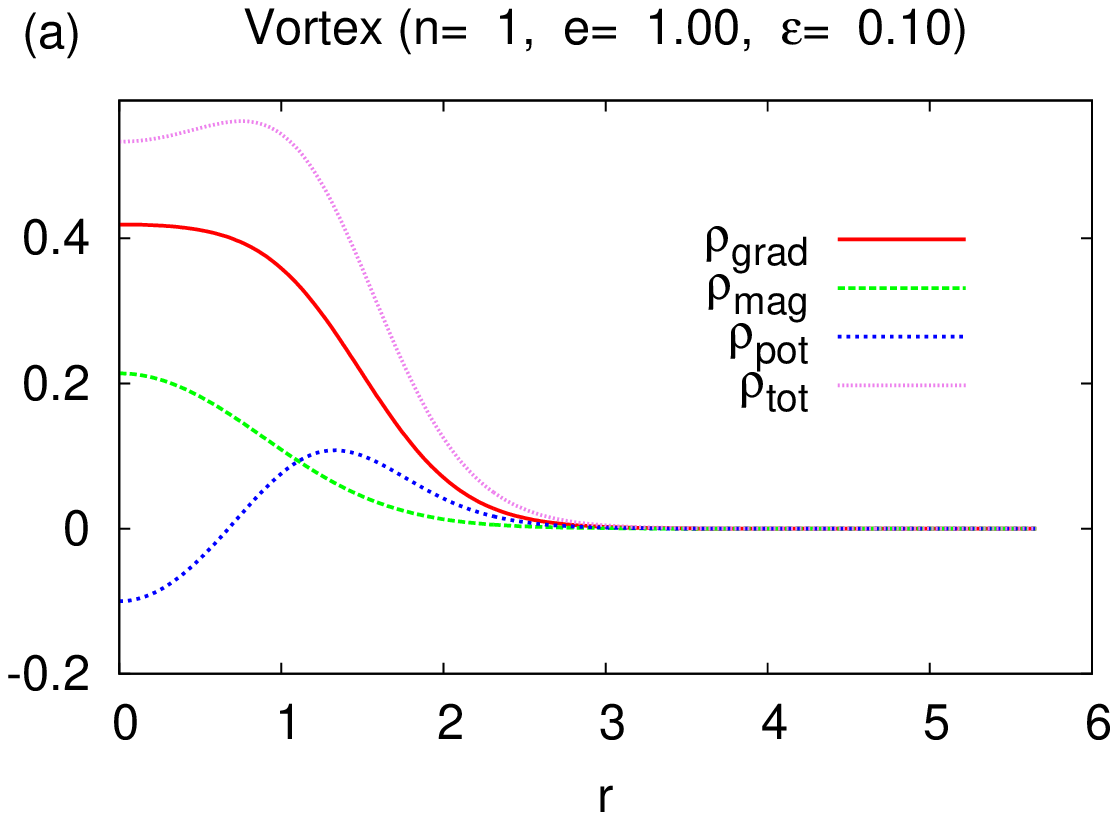}
\qquad
\includegraphics[width=7cm]{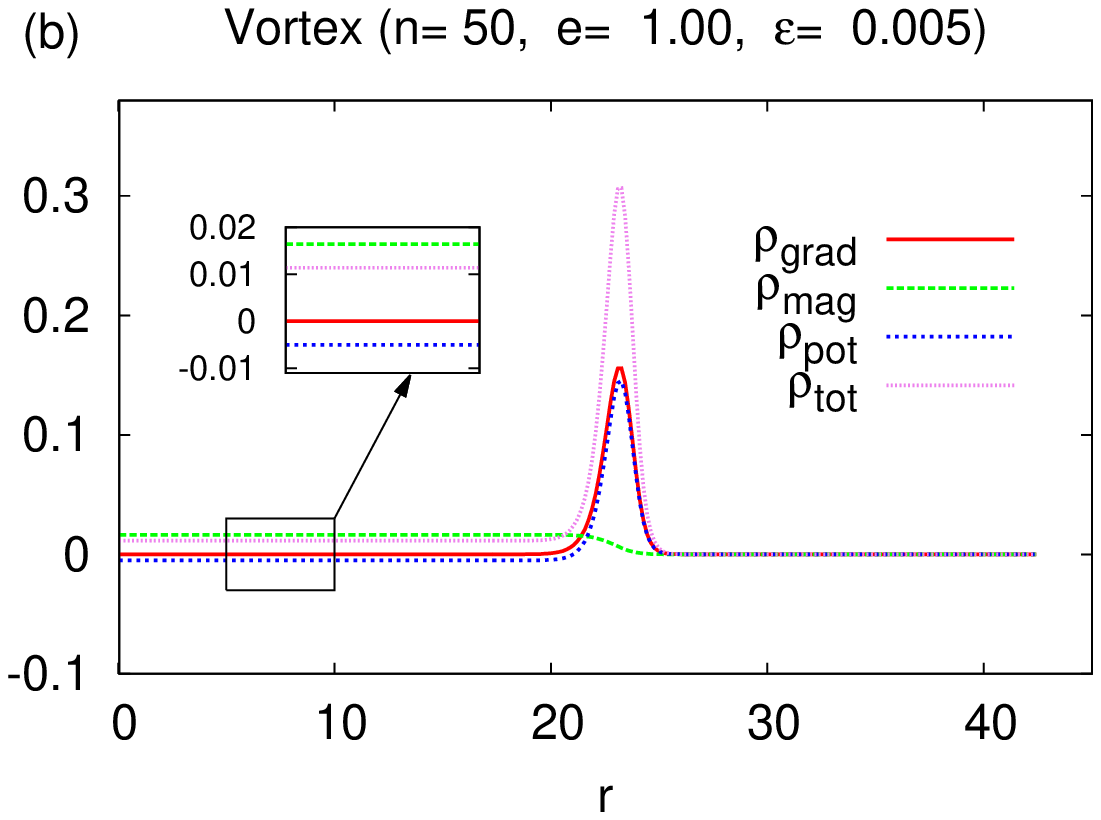}
\end{center}
\caption{(Color online) Scalar field gradient energy density $\rho_{\rm grad}$,
magnetic field energy density $\rho_{\rm mag}$,
potential energy density $\rho_{\rm pot}$,
and total energy density $\rho_{\rm tot}$ for (a) thick-wall and (b) thin-wall vortices.}
\label{fig:fig03}
\end{figure}

The various contributions to the energy density as well as the total energy density are shown for these solutions in Fig.~\ref{fig:fig03}; all of these vanish as $r\to\infty$, as expected. Note that in both cases the potential energy density is negative as $r\to0$; it then rises to a maximum and returns to zero, as expected given the profile of $f(r)$ and the form of the potential (see Fig.~\ref{fig:fig01}). The total energy of the thick-wall vortex is 5.38, which will be compared with that of an ansatz we will use in the next section. That of the thin-wall vortex for the parameters of Fig.~\ref{fig:fig02}b is  92.5.

\begin{figure}[ht]
\begin{center}
\includegraphics[width=7cm]{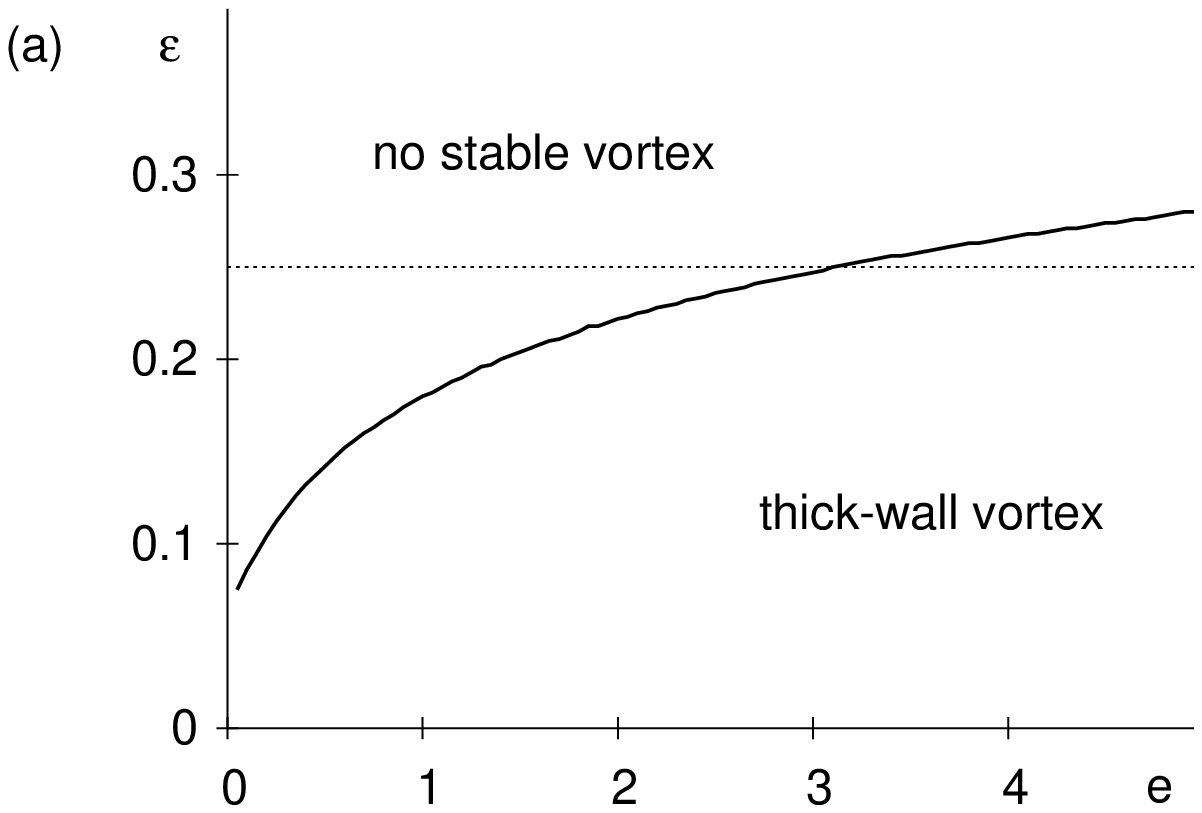}
\qquad
\includegraphics[width=7cm]{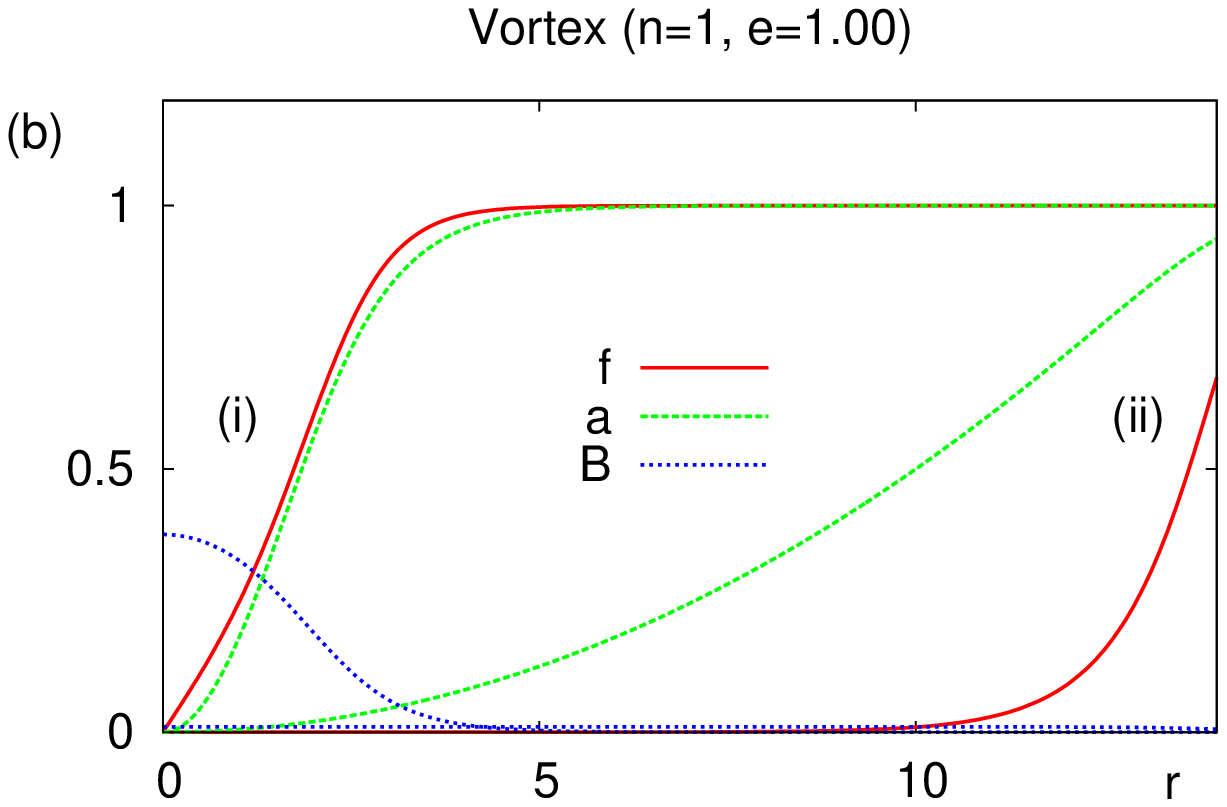}
\end{center}
\caption{(Color online) (a) Scan of parameter space for $n=1$, illustrating what can be described as a sort of phase transition dividing parameter space into two regions, one where classically stable thick-wall vortices exist and one where they do not exist. Curved solid line is actual phase transition calculated numerically. Horizontal dotted line is that corresponding to piecewise ansatz (see (\ref{eq:narrowenergy1})). (b) Numerical solutions found for (i) $\eps=0.179$, in the stable region, and (ii) $\eps=0.180$, in the unstable region. Although very close to the phase transition, the stable solution is qualitatively very similar to that displayed in Fig.~\ref{fig:fig02}a. The unstable solution displayed is in fact an artifact of the numerical algorithm, arising because the boundary condition imposed at the maximal radius presupposes $f\simeq1$ (see the discussion following (\ref{boundary1},\ref{boundary2})). For instance, if the range of integration (that is, the maximum radius) is increased, the numerical solution expands along with it. No such finite-size effects occur for the stable solution. }
\label{fig:scan}
\end{figure}

Let us finish this section with a few comments regarding vortices with $n=1$. As $\eps$ varies, the potential changes in a way that has a dramatic effect on the vortices; this effect can be described as a sort of phase transition in that below a certain critical value of $\eps$ (which depends on $e$), thick-wall vortices exist while above it no stable vortices are found. (It appears that thin-wall vortices are not seen for this potential for $n=1$.) This phase transition is not a complete surprise: for $\eps<0$, $\phi=0$ has greater potential energy than $\phi=1$, so the vortex is certainly stable. For reasons outlined above, this is no longer clear if $\eps>0$, and indeed as $\eps$ goes from below 1 to above, $\phi=1$ goes from being a local minimum to a local maximum, and solutions of the form we are looking for certainly do not exist. This is borne out by a scan of parameter space, (see Fig.~\ref{fig:scan}a) which shows that the stable thick-wall vortex no longer exists well below $\eps=1$. There is also a relatively mild dependence on $e$, although $e\to0$ is a delicate limit since setting $e=0$ decouples the gauge and scalar fields, and the vortex becomes a global topological defect.

There is remarkably little variation in the form of the stable vortex as the parameters vary. Of course the details depend on the parameters, but generally the stable vortices look much like that displayed in Fig.~\ref{fig:fig02}a, even when very close to the phase transition, as illustrated in Fig.~\ref{fig:scan}b.

Although thin-wall vortices (for $n=1$) do not exist in the above model, unstable thin-wall configurations of large radius can be constructed, so one can imagine a tunneling event from the thick-wall solutions above to an unstable thin-wall configuration of the same energy which would then expand rapidly. In the next section we will first discuss the quantum mechanical tunneling of $n=1$ vortices to such unstable thin-wall configurations.  Subsequently, we will consider the tunneling of large-$n$ thin-wall vortices, where the thin-wall nature simplifies the analysis.
\vfill\eject

\section{Fate of the false vortex via tunneling \label{sec-fate}}

We wish in this section to study the quantum mechanical decay of false vortices in the model discussed above. There are two reasons for doing so. First, it is intrinsically interesting to determine the vortex lifetime. Second, in a cosmological model where the universe is in a false vacuum with $|\phi |=1$, it will eventually tunnel to the true vacuum with a rate which can be calculated following the standard method presented in \cite{Coleman}. But the universe would generically contain vortices, and it is interesting and potentially important to see what effect a gas of false vortices would have on the tunneling rate. One might imagine that, given that the core of the vortex is already in the true vacuum, the presence of vortices could cause the universe to tunnel more rapidly than it would in the absence of vortices. We will discuss these matters in this section, beginning with the case of thick-wall vortices. 
\subsection{Thick-wall ansatz ($n=1$)}\label{subsec-thick}

The action of the ansatz (\ref{ansatz}) with $n=1$ is
\[
S=\int dt \, (T-E)
\]
where $T$ is the kinetic energy
\beq
T = 2\pi \int_{0}^{\infty} dr\,r \left( \dot{f}^{2} + \frac{\dot{a}^{2}}{2e^{2}r^{2}} \right)
\label{eq:kinetic1}
\eeq
where $E$ is the energy of a static configuration
\beq
E = 2\pi \int_{0}^{\infty} dr\,r \left( f'^{2}+\frac{(1-a)^{2}}{r^{2}}f^{2} + \frac{a'^{2}}{2e^{2} r^{2}} + (f^2-\eps) (f^2-1)^2 \right).
\label{eq:energy1}
\eeq
In this section, for illustrative purposes we will use parameter values $e=1$, $\eps=0.1$.

In principle, we would like to find the instanton (or bounce), which in our case is the solution to the Euclidean field equations which tends towards the vortex as Euclidean time $\tau\to-\infty$, reaches a turnaround point at $\tau=0$, and returns to the vortex as $\tau\to+\infty$. This is a daunting task, and we will instead analyze a simplified problem replacing the full space of field configurations by a one-parameter family of configurations, parameterized by the radius of the vortex. This reduces the problem to a one-dimensional tunneling problem. In the semi-classical approximation, the tunneling rate of this one-dimensional problem is proportional to $e^{-S_E}$, where $S_E$ is the action of the solution of the Euclidean equation of motion with the appropriate boundary conditions. Since we have not solved the full field equations, the actual bounce action will be lower, so the tunneling rate determined from the one-dimensional tunneling problem will be a lower bound on the true semi-classical tunneling rate.

\begin{figure}[ht]
\begin{center}
\includegraphics[width=7cm]{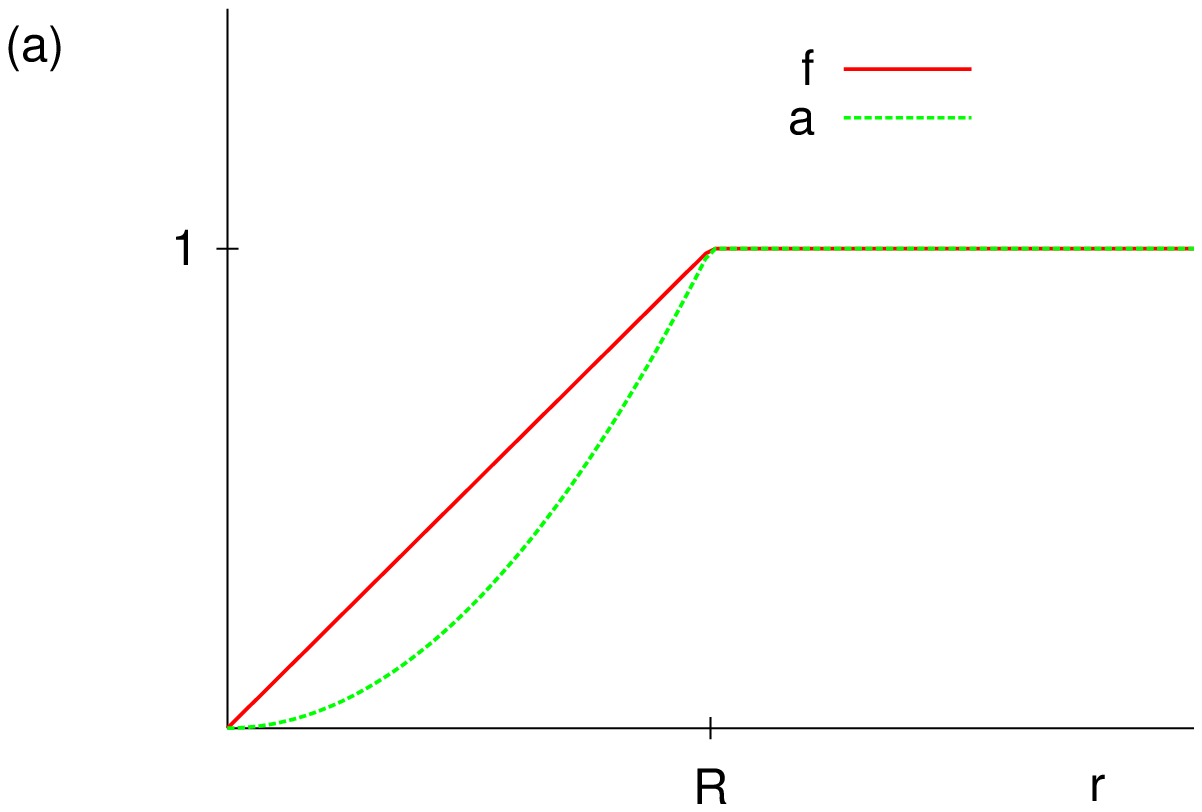}
\qquad
\includegraphics[width=7cm]{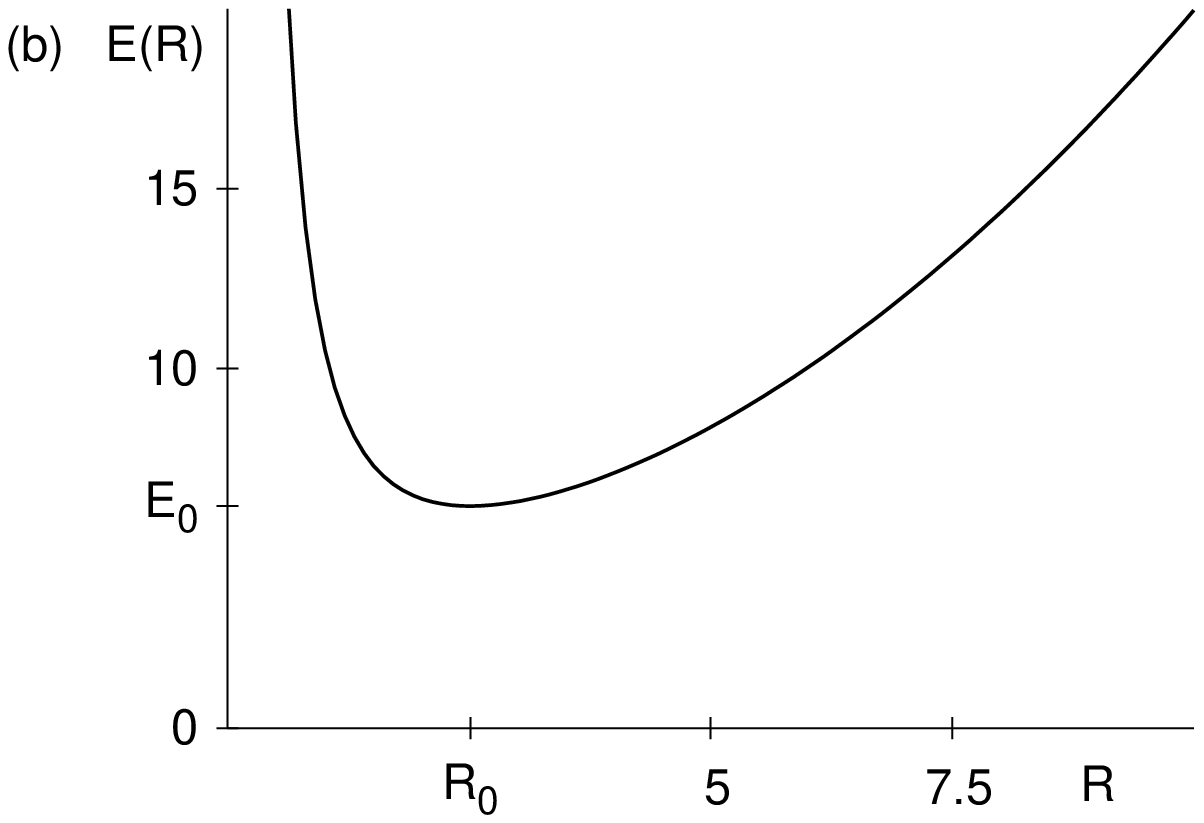}
\end{center}
\caption{(Color online) (a) Piecewise vortex ansatz. (b) Vortex energy E(R) for $e=1$, $\eps=0.1$.
\label{fig:thickwall1}}
\end{figure}

We begin by determining the minimum-energy configuration within a family of configurations representing a vortex of width $R$, treating $R$ as a variational parameter. This family is given by (see Fig.~\ref{fig:thickwall1}a)
\beq
f(r)=\left\{\begin{array}{ccc}r/R&&r<R\\ 1&&r>R\end{array}\right.,
\qquad\qquad
a(r)=\left\{\begin{array}{ccc}(r/R)^2&&r<R\\ 1&&r>R\end{array}\right. .
\label{eq:thickwall1}
\eeq
For any $R$, this configuration has the correct asymptotic form both as $r\to0$ and as $r\to\infty$. The field $a$ describes a uniform magnetic field for $r<R$ with unit flux.
The energy as a function of $R$ is
\bea
E(R)&=&\frac{4\pi}{3} + \frac{2\pi}{e^2R^2}
+ \frac{\pi R^2(1-4\eps)}{12} \nonumber\\
&=&\frac{4\pi}{3} + \frac{2\pi}{R^2} + \frac{\pi R^2}{20}.
\label{eq:narrowenergy1}
\eea
where the latter expression is for $e=1$, $\eps=0.1$. $E(R)$
has a minimum for any $e$ as long as $\eps<1/4$, which bears at least some resemblance to the scan of parameter space for the exact numerical solution displayed in Fig.~\ref{fig:scan}. Let its minimum value be $E_0$ at $R=R_0$. For the values of the parameters considered in this section, $E_0=4\pi/3+2\pi/\sqrt{10}\simeq 6.18$ and $R_0=40^{1/4}\simeq 2.51$. The former is somewhat higher than the actual minimum value 5.38 given above, as anticipated; the latter is in excellent qualitative agreement with Fig.~\ref{fig:fig02}a.

\begin{figure}[ht]
\begin{center}
\includegraphics[width=8cm]{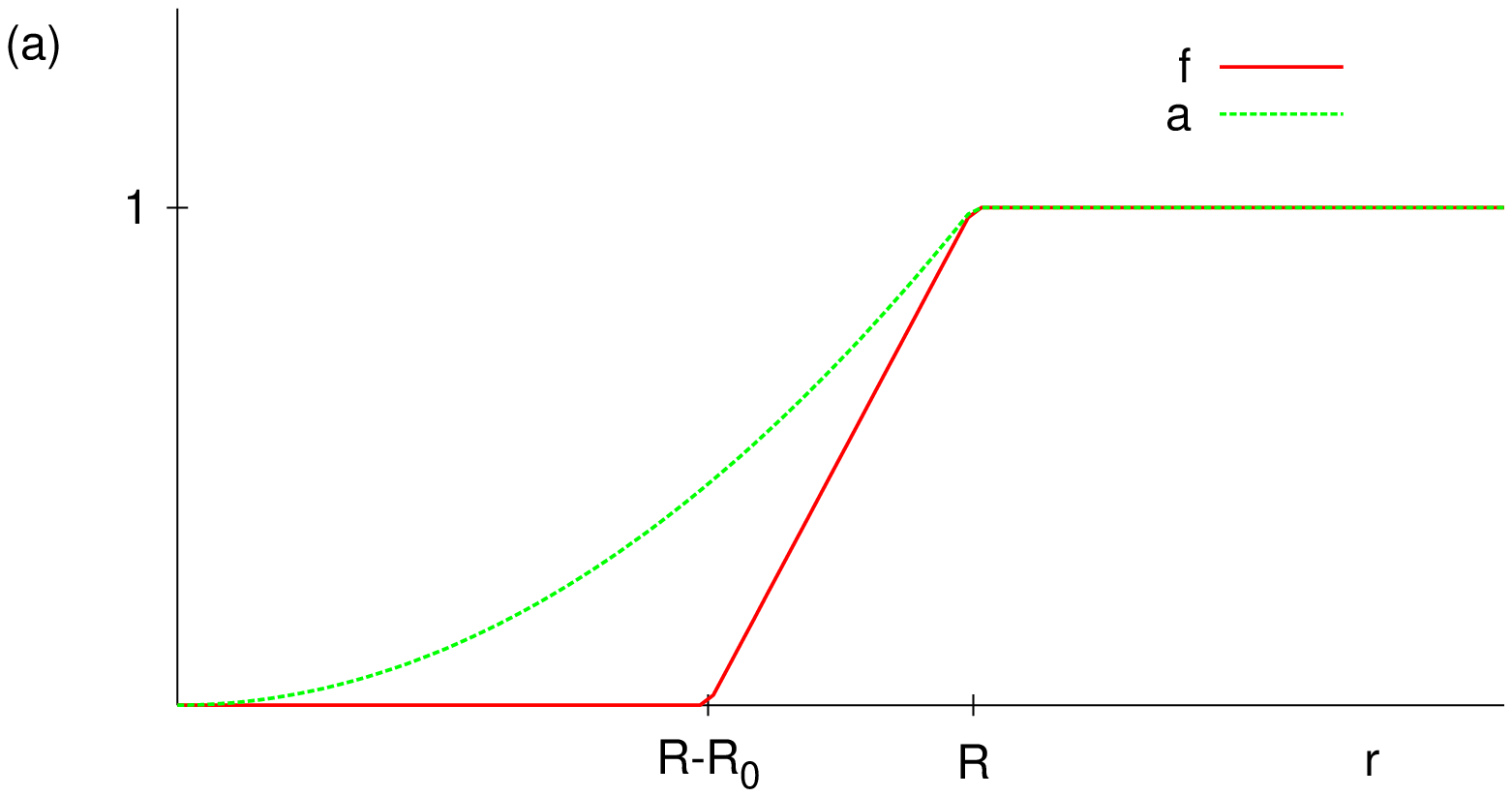}
\qquad
\includegraphics[width=7cm]{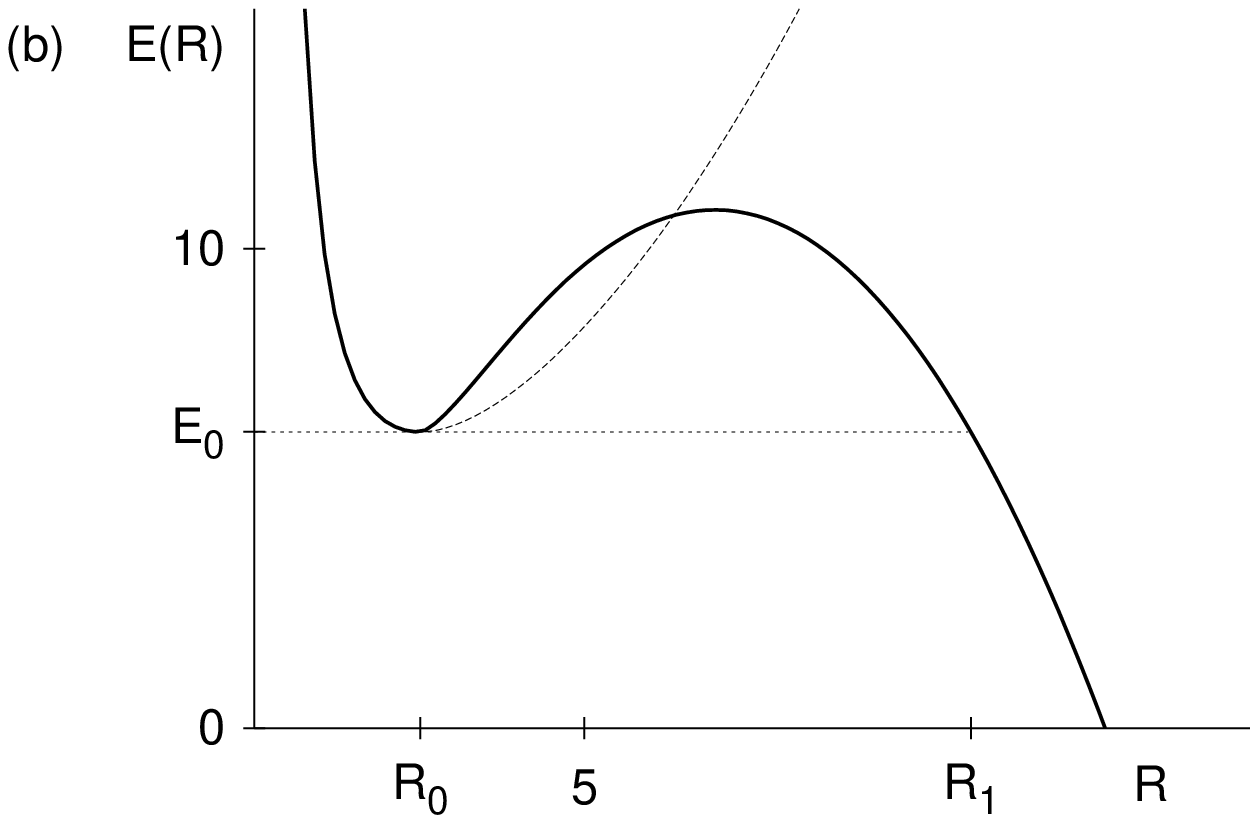}
\end{center}
\caption{(Color online) (a) Ansatz for vortex configurations with $R>R_0$. (b) Vortex energy $E(R)$ for $e=1$, $\eps=0.1$, using (\ref{eq:narrowenergy1}) for $r<R_0$ and (\ref{eq:wideenergy1}) for $r>R_0$. Dashed curve represents the continuation of (\ref{eq:narrowenergy1}) outside its domain of validity for the tunneling process.
\label{fig:thickwall2}}
\end{figure}

We will assume the tunneling proceeds within a space of configurations parameterized by $R(t)$, defined as follows. For $R<R_0$, the configuration is a ``squeezed" vortex described by (\ref{eq:thickwall1}). For $R>R_0$, the vortex will be assumed to have the following profile (see Fig.~\ref{fig:thickwall2}a):
\beq
f(r)=\left\{\begin{array}{ccc}0&&r<R-R_0\\
\frac{r-(R-R_0)}{R_0}&&R-R_0<r<R\\
1&&r>R\end{array}\right. ,
\qquad\qquad                                         
a(r)=\left\{\begin{array}{ccc}(r/R)^2&&r<R\\ 1&&r>R\end{array}\right. .
\label{eq:thickwall2}
\eeq
The energy of this configuration is given by substituting (\ref{eq:thickwall2}) into (\ref{eq:energy1}), which yields
\bea
E(R)&=&\frac{2\pi}{e^2R^2} + 2\pi\left(\frac{R}{R_0}-1\right)^2\log\frac{R}{R-R_0}\nonumber\\
&&+\frac{\pi}{30R^4}\left(60R^4-20R^3R_0-5R^2{R_0}^2+6R{R_0}^3-{R_0}^4\right)\nonumber\\
&&\quad+\frac{\pi {R_0}}{420}\left(64R-29R_0\right)
-\frac{\pi \eps}{15}\left(15R^2-14R{R_0}+4{R_0}^2\right)\nonumber\\
&=&\frac{2\pi}{R^2} + 2\pi\left(\frac{R}{R_0}-1\right)^2\log\frac{R}{R-R_0}\nonumber\\
&&+\frac{\pi}{30R^4}\left(60R^4-20R^3R_0-5R^2{R_0}^2+6R{R_0}^3-{R_0}^4\right)\nonumber\\
&&\quad-\frac{\pi}{700}(70R^2-43R R_0+67{R_0}^2),
\label{eq:wideenergy1}
\eea
where the latter expression is for $e=1$, $\eps=0.1$.

Eqs. (\ref{eq:narrowenergy1}) (for $R<R_0$) and (\ref{eq:wideenergy1}) (for $R>R_0$) give the potential for the one-dimensional problem, which is displayed in Fig.~\ref{fig:thickwall2}b. It has the expected form, with a local minimum at $R=R_0$ separated by a barrier from the region corresponding to a wide thin-wall vortex whose energy is unbounded from below as $R\to\infty$. $R_1$ is the radius at which the thin-wall vortex energy equals that of the vortex solution; for the values of the parameters considered in this section, $R_1=10.86$.

The tunneling process corresponds to a quantum transition from $R_0$ through the classically-forbidden region to $R_1$. This transition is mediated by an instanton or bounce solution of the Euclidean equations of motion.  We will find an approximation to the exact bounce by analyzing the (Euclidean) dynamical equations for $R(\tau)$, where $\tau$ is the Euclidean time.  

The tunneling problem is defined by a potential which is $-E(R)$ (shifted so that the energy of the one-dimensional problem is zero) combined with a kinetic energy term which is determined by substituting (\ref{eq:thickwall2}) (with $R\to R(\tau)$) into (\ref{eq:kinetic1}), the dot now being interpreted as a derivative with respect to $\tau$. This yields the following expression for the Euclidean action of an arbitrary path $R(t)$:
\beq
S_E[R(\tau)]=\int d\tau \left( B(R){\dot R}^2 + (E(R)-E_0) \right),
\label{eq:action2}
\eeq
with $E(R)$ as given in (\ref{eq:wideenergy1}) and
\beq
B(R)=\pi\left(\frac{2R}{R_0}-1+\frac{1}{e^2R^2}\right).
\label{eq:B}
\eeq

The action (\ref{eq:action2}) can be interpreted as that of a particle with a position-dependent mass moving in the potential $E_0-E(R)$. We are interested in the bounce, the solution for which the particle starts at rest at $R=R_0$ as $\tau\to-\infty$, rolls to $R_1$ and then returns to $R_0$ as $\tau\to+\infty$. In Euclidean spacetime, the bounce corresponds to a vortex whose radius varies in time, starting at $R_0$, increasing to $R_1$ and then returning to $R_0$. Of course, this is a Euclidean motion, not an actual physical motion. In real spacetime, the most that can be said is that a quantum fluctuation takes the vortex to the larger radius $R_1$, after which it expands normally according to the Minkowski equations of motion (essentially, rolling down the slope at $R=R_1$ in Fig.~\ref{fig:thickwall2}b towards $R=\infty$). Since both the mass and the potential are independent of time, the resulting Euler-Lagrange equation has a first integral, which is
\beq
B(R) {\dot R}^2-E(R)+E_0=0,
\eeq
where we have used the fact that the particle starts at rest at $R_0$. The action for the classical motion starting at $R_0$ and ending at $R_1$ is
\beq
\label{eq:thickwallbounceaction}
S_B^{\rm thick}=2\int_{R_0}^{R_1} dR \sqrt{B(R)(E(R)-E_0)};
\eeq
the factor 2 is because the bounce is a ``round trip" between $R_0$ and $R_1$. The integral cannot be evaluated, except by numerical integration. The result is not terribly illuminating, and rather than pursue this we will consider a different case where the analysis can be pushed further in the next subsection.

\subsection{Thin-wall vortices ($n\gg1$)}\label{subsec-thin}

As we saw earlier, thin-wall vortices exist if $n$ is sufficiently large. In this case, we can imagine that the bounce is a sequence of thin-wall vortices in a potential much like that depicted in Fig.~\ref{fig:thickwall2}b (the details of course will be different). The fact that the vortex is of thin-wall type allows us to compute explicitly the Euclidean action within the thin-wall approximation.

It is intuitively clear that thin-wall vortices will occur if $n$ is sufficiently large, roughly because the vortex size will increase with magnetic flux, while the spatial scale over which the scalar field varies is on the order of its Compton wavelength which is independent of $n$. Thus, the transition zone from the interior of the vortex, where $f\simeq 0$, to the exterior, where $f\simeq 1$, occurs over some length scale $\delta\sim 1$ while $R\gg 1$, in agreement with the numerical solution depicted in Fig.~\ref{fig:fig02}b. The existence of thin-wall vortices has also been shown explicitly and studied from various points of view in \cite{bol1,bol2,bol3,sut1}.

We must reintroduce $n$ into the action, giving
\[
S=\int dt\, (T-E)
\]
where the kinetic term is
\beq
T = 2\pi \int_{0}^{\infty} dr\,r \left( \dot{f}^{2} + \frac{n^2\dot{a}^{2}}{2e^{2}r^{2}} \right)
\label{eq:action3}
\eeq
and the energy of a static configuration is now
\beq
E = 2\pi \int_{0}^{\infty} dr\,r \left( f'^{2}+\frac{n^2(1-a)^{2}}{r^{2}}f^{2} + \frac{n^2a'^{2}}{2e^{2} r^{2}} + (f^2-\eps) (f^2-1)^2 \right).
\label{eq:energy3}
\eeq
Let us first determine the energy as a function of $R$ of a static thin-wall configuration. We can divide the energy integral (\ref{eq:energy3}) into three regions:
\beq
E(R)=E_{\rm int} + E_{\rm wall} + E_{\rm ext}.
\label{eq:3regions}
\eeq

In the interior region ($r<R-\delta/2$), we assume that the fields are $f(r)=0,\ a(r)=(r/R)^2$. Then only the third and fourth terms of (\ref{eq:energy3}) contribute and we find
\beq
E_{\rm int} = \frac{2\pi n^2}{e^2 R^2}-\eps\pi R^2
\label{eq:Eint}
\eeq
where we have dropped corrections smaller by a factor $\delta/R$.

Inside the wall ($R-\delta/2<r<R+\delta/2$), we can replace factors $r$ in the integral by $R$, to leading order. In the second and third terms of (\ref{eq:energy3}) we can assume $1-a\sim 1/R$, $a'\sim 1/R$ and $f\simeq 1$, so (assuming $e\sim 1$) these terms are of order $n^2/R^3$. Since we expect $R\gg1$, these terms are small compared to the first term of (\ref{eq:Eint}) so they can safely be dropped to leading order. For the first and fourth terms of (\ref{eq:energy3}), the equation of motion for $f$ is
\beq
\label{eq:f}
f'' + \frac{f'}{r} - \frac{n^2}{r^2} (1-a)^2 f -(f^2-1)(3f^2-(1+2\eps))f = 0.
\eeq
The second and third terms of this equation can be dropped since $r\gg1$ in this region. Multiplying by $f'$, we can integrate the equation, giving
\beq
f'^2=(f^2-\eps)(f^2-1)^2.
\eeq
Thus,
\bea
\label{eq:Ewall}
E_{\rm wall} &=& 4\pi R \int_{R-\delta/2}^{R+\delta/2}dr\, f'^2
= 4\pi R \int_{R-\delta/2}^{R+\delta/2} dr\, f'\sqrt{(f^2-1)^2(f^2-\eps)}\nonumber \\
&=& 4\pi R \int_0^1 df \sqrt{(f^2-1)^2(f^2-\eps)}
\eea
To leading order, we can put $\eps=0$, giving $E_{\rm wall}=\pi R$.

\begin{figure}[ht]
\begin{center}
\input{Ehat-take2.tex}
\end{center}
\caption{The rescaled energy $\hE(\hR)$ with $\hat\eps=0.108$. Numerical values for the three parameters shown are: $\hE_0=1.38$, ${\hR}_0=1.09$, ${\hR}_1=7.61$.
} \label{fig:Ehat}
\end{figure}
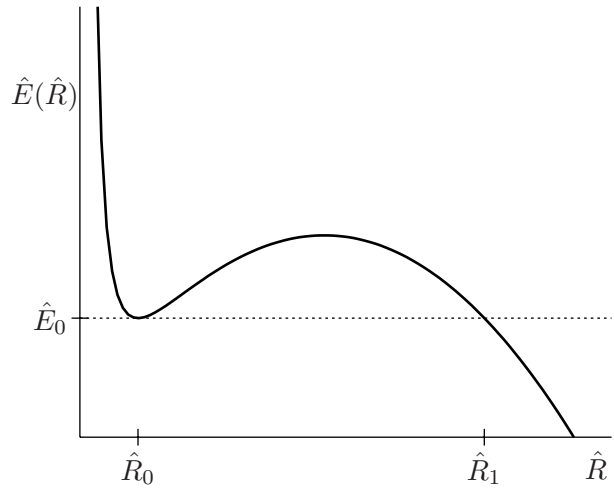

In the exterior region ($r>R+\delta/2$),  we assume $f(r)=a(r)=1$, and we find $E_{\rm ext}=0$. Summing the three contributions, we find
\beq
\label{eq:EofR}
E(R)= \frac{2\pi n^2}{e^2 R^2}+\pi R-\eps\pi R^2,
\eeq
Finding the expected minimum of this function involves solving a fourth-order polynomial equation which cannot be done exactly. It is useful to define $\hR=(2n/e)^{-2/3}R$, $\hE(\hR)=(2n/e)^{-2/3} E(R)/\pi$ and $\hat\eps=(2n/e)^{2/3}\eps$. In terms of these new variables,
\beq
\label{eq:Ehat}
\hE(\hR)=\frac{1}{2{\hR}^2} + \hR - \hat \eps {\hR}^2,
\eeq
involving only one parameter $\hat\eps$ which we assume is small. This function is displayed in Fig.~\ref{fig:Ehat} for the parameters used in Fig.~\ref{fig:fig02}b (for which $\hat\eps=0.108$).

$\hE(\hR)$ displays the same qualitative features for any positive $\hat\eps$ smaller than a critical value, which turns out to be ${\hat\eps}_c=3/2^{11/3}\simeq0.24$. At that value, the points $\hR_0$ and $\hR_1$ coalesce, forming an inflection point at $(\hR,\hE)=(\hR_c,\hE_c)=(2^{4/3},3/2^{4/3})$, and the tunneling barrier disappears. If $\hat\eps>{\hat\eps}_c$, $\hE(\hR)$ is a monotonic decreasing function: there is no classically stable vortex. We will return to the near-critical case $\hat\eps\leq{\hat\eps}_c$ in subsection \ref{subsec-dissociation}.

We can calculate $\hR_0$ and $\hE_0$ perturbatively in $\hat\eps\ll1$; to lowest order we can simply drop the third term of (\ref{eq:Ehat}), and we find $\hR_0=1,\ \hE_0=3/2$, with corrections (positive and negative, respectively) of order $\hat\eps$. The larger radius $\hR_1$ does not exist for $\hat\eps=0$, but it can be written as a Laurent series; it is $\hR_1=1/\hat\eps$ with a (negative) correction of order 1. These corrections are easily calculated and are in good agreement with the values given in Fig.~\ref{fig:Ehat}. In terms of the original variables, we find
\beq
\label{eq:origvar}
R_0=\left(\frac{2n}{e}\right)^{2/3}, \quad
E_0=\frac{3\pi}{2}\left(\frac{2n}{e}\right)^{2/3},\quad
R_1=\frac{1}{\eps}
\eeq
with corrections smaller by a factor $\hat\eps=(2n/e)^{2/3}\eps$. $R_0$ and $E_0$ are in good agreement with the size and energy of the thin-wall vortex found numerically earlier (see Figs.~\ref{fig:fig02}b, \ref{fig:fig03}b).

The form of $E(R)$ indicates that bounce solutions exist. In order to compute their Euclidean action, we need to determine the kinetic term (\ref{eq:action3}). Once again we can consider three regions (interior, wall and exterior). In the interior, $f=0$ while $a=(r/R(t))^2$, so $\dot a = -2r^2\dot R/R^3$, and we find
\beq
\label{eq:Tint}
T_{\rm int} = \frac{\pi n^2}{e^2}\frac{{\dot R}^2}{R^2}.
\eeq
Inside the wall, we can assume that $f(r,t)$ is a function of $r-R(t)$, that is, the time dependence of $f(r,t)$ is due only to translation by $R(t)$, the position of the wall. Then
\[
\frac{\partial f}{\partial t}=-f'(r)\dot R(t)
\]
and the first term of (\ref{eq:action3}) is
\[
2\pi\int_{R-\delta/2}^{R+\delta/2}dr\,f'^2 {\dot R}^2 = \frac{\pi R}{2}{\dot R}^2,
\]
where we have used the fact that the integral has already been evaluated in (\ref{eq:Ewall}). As was the case with the evaluation of the energy, the second term of (\ref{eq:action3}) is smaller than (\ref{eq:Tint}) by a factor $(\delta/R)$ so it can be dropped, and
\[
T_{\rm wall} = \frac{\pi R}{2}{\dot R}^2.
\]
Outside the wall both $f$ and $a$ are constant and the contribution to (\ref{eq:action3}) vanishes, so the kinetic term is
\beq
\label{eq:kineticenergy}
T = T_{\rm int} + T_{\rm wall} = \left( \frac{\pi n^2}{e^2R^2} + \frac{\pi R}{2} \right) {\dot R}^2.
\eeq

We can now write the Euclidean action:
\[
S_E = \int dt\, \left\{\left( \frac{\pi n^2}{e^2R^2} + \frac{\pi R}{2} \right) {\dot R}^2
+ \frac{2\pi n^2}{e^2 R^2}+\pi R-\eps\pi R^2 - E_0\right\}.
\]
We are now in a position to write an expression for the bounce action, following the procedure used to obtain (\ref{eq:thickwallbounceaction}). The result is
\bea
\label{eq:thinbounceaction1}
S_B^{\rm thin}&=&2\int_{R_0}^{R_1} dR \sqrt{
\left( \frac{\pi n^2}{e^2R^2} + \frac{\pi R}{2} \right)
\left(\frac{2\pi n^2}{e^2 R^2}+\pi R-\eps\pi R^2 - E_0\right)}\nonumber\\
&=&2\pi\left(\frac{2n}{e}\right)^{4/3}
\int_{{\hR}_0}^{{\hR}_1}\frac{d\hR}{{\hR}^2}
\sqrt{
\left(\frac14+\frac{{\hR}^3}{2}\right)
\left(\frac12 - \hE_0{\hR}^2 + {\hR}^3 - \hat\eps{\hR}^4\right)
}
\eea
where we have gone to the hatted variables defined earlier.

Note that the last factor is the energy function $\hE(\hR)-\hE_0$ multiplied by ${\hR}^2$. This function, a quartic polynomial, has four real roots, two of which are ${\hR}_0$ and one of which is ${\hR}_1$. The final root, which we will call ${\hR}_2$, can be determined as a power series in $\hat\eps$; we find ${\hR}_2=-.5+O(\hat\eps)$. Thus we can write
\beq
\label{eq:thinbounceaction2}
S_B^{\rm thin}=2\pi\sqrt{\frac{\hat\eps}{2}}\left(\frac{2n}{e}\right)^{4/3}
\int_{{\hR}_0}^{{\hR}_1}\frac{d\hR}{{\hR}^2}
\left({\hR}-{\hR}_0\right)\sqrt{
\left({\hR}^3+\frac12\right)
\left({\hR}_1-{\hR}\right)\left({\hR}-{\hR}_2\right)
}.
\eeq
For small $\hat\eps$, we can make the approximations ${\hR}_0=1$, ${\hR}_1=1/\hat\eps$, ${\hR}_2=-.5$, $\hE_0=3/2$, giving
\beq
\label{eq:thinbounceaction3}
S_B^{\rm thin}=2\pi\sqrt{\frac{\hat\eps}{2}}\left(\frac{2n}{e}\right)^{4/3}
\int_{1}^{1/{\hat \eps}}\frac{d\hR}{{\hR}^2}
\left({\hR}-1\right)\sqrt{
\left({\hR}^3+\frac12\right)
\left(\frac{1}{\hat\eps}-{\hR}\right)\left({\hR}+\frac12\right)
}.
\eeq
Finally, since the integral is dominated by $\hR \gg 1$, we can drop the factors $1/2$ (an approximation whose validity can easily be verified), after which the integral can be evaluated exactly, giving
\bea
\label{eq:thinbounceaction4}
S_B^{\rm thin}&=&2\pi\sqrt{\frac{\hat\eps}{2}}\left(\frac{2n}{e}\right)^{4/3}
\int_{1}^{1/{\hat \eps}}d\hR\,
\left({\hR}-1\right)\sqrt{ \left(\frac{1}{\hat\eps}-{\hR}\right) }\nonumber\\
&=&
2\pi\sqrt{\frac{\hat\eps}{2}}\left(\frac{2n}{e}\right)^{4/3}
\frac{4}{15}\left(\frac{1}{\hat\eps}-1\right)^{5/2}
\simeq
\frac{4\sqrt{2}\pi}{15}\frac{1}{\eps^2}
\eea
where in the last step we have made the approximation $(1/\hat\eps)-1\simeq1/\hat\eps$ and we have also returned to the original variables. Interestingly, the action is independent of $n$ (aside from the fact that \eqref{eq:thinbounceaction4} was derived for thin-wall vortices, an approximation that is valid only for $n\gg1$).

Recall that $S_B^{\rm thin}$ is an upper bound to the bounce action; thus, it gives a lower bound on the decay rate for the vortex, which is:
\beq
\label{eq:Gathin}
\Ga^{\rm thin}=A^{\rm thin} e^{-S_B^{\rm thin}}.
\eeq
The coefficient $A^{\rm thin}$ comes from the determinant arising in the saddle-point evaluation of the path integral, as discussed in \cite{Coleman,calcol77}.  This determinant factor must exclude the integration over the zero modes.  The only zero mode of the vortex is because of time translation invariance.  The position of the vortex is fixed once and for all.  Thus the integration over the direction of the time translation zero mode in the determinant  is removed and instead the time at which the bounce occurs is integrated over.  This change of variables gives rise to a Jacobian factor which is evaluated in \cite{Coleman} and yields the decay rate of the vortex 
\beq
\label{eq:Gathin1}
\Ga^{\rm thin}=A^{\rm thin} \left(\frac{S_B^{\rm thin}}{2\pi}\right)^{1/2}  e^{-S_B^{\rm thin}}
\eeq
with a minor abuse of notation, as $A^{\rm thin} $ now is the determinant excluding the zero mode.

As stated earlier, we are interested in comparing the decay rate of a gas of vortices, each of which decays with rate \eqref{eq:Gathin1}, with that of the ordinary (translation-invariant) vacuum. For the latter, we imagine that the universe is in a false vacuum with $\phi=1$ (up to an irrelevant position-independent phase). The decay rate per unit volume is calculated using the method of \cite{Coleman}. The bounce is the path in configuration space of least action, and we assume that $\phi$ is always real (in what follows we write $\phi=f$) and that the gauge fields are not excited. Thus, we work with the Lagrangian
\[
{\cal L} = (\partial_\mu f)^2 - V(f),
\]
with $V$ as in (\ref{potential}). Furthermore, we assume that the bounce is a function only of the Euclidean radial coordinate $\rho=\sqrt{\tau^2+{\bf x}^2}$. The Euclidean action and equation of motion are then
\[
S_B^{\rm vac}=4\pi \int_0^\infty d\rho\,\rho^2\,\left( f'^2 + V(f) \right),
\qquad
f'' + \frac2\rho f' = V'(f).
\]
If we view $f$ as the position of a particle on a half-line and $\rho$ as a time coordinate, this equation describes the particle moving in a potential $-V$ with a time-dependent friction term. The bounce solution has $f(\rho=0)\simeq 0$ and $f'(\rho=0)=0$, the exact starting value being that for which $f(\rho)\to1$ as $\rho\to\infty$.
As demonstrated in \cite{Coleman}, if we assume that $\eps\ll1$, then the bounce will be of thin-wall type, with $f$ staying very near the false vacuum $f=0$ for a long time, then making a rapid transition to near $f=1$, asymptotically approaching that value as $\rho\to\infty$. Since the transition occurs for $\rho\gg1$, the friction term can be neglected and the potential can be taken to be that with $\eps=0$. The equation of motion is then that of a soliton much like the kink of $\phi^4$ theory:
\beq
\label{eq:fprime}
f'=\sqrt{\left.V(f)\right|_{\eps=0}}=f(1-f^2).
\eeq
It is easily integrated to obtain the explicit profile (call it $f_K(\rho)$), but in fact this is not necessary; all we need is the action, which is
\[
S_B^{\rm vac} = 4\pi\int_0^\infty d\rho\,\rho^2\left({f_K'}(\rho)^2 + V(f_K)\right)
= \frac43\pi R^3 (-\eps) + 4\pi R^2 \int_0^\infty d\rho\,{f_K'}(\rho)^2.
\]
Eq.~(\ref{eq:fprime}) enables us to write the latter integral
\[
\int_0^\infty d\rho\,{f_K'}(\rho)^2=\int_0^1df\,f(1-f^2)=\frac14,
\]
so
\[
S_B^{\rm vac} = \pi \left( R^2-\frac43 \eps R^3\right).
\]
Minimizing with respect to $R$ gives us the radius of the thin-wall bounce as well as its action:
\beq
\label{eq:SBvac}
R=\frac{1}{2\eps},
\qquad\qquad
S_B^{\rm vac} = \frac{\pi}{12\eps^2}.
\eeq

The thin wall bubble admits three zero modes corresponding to space-time translation invariance. Thus the determinant factor must exclude these modes, while we must integrate over the space-time position of the bubble.  This gives rise to a factor of $\left(S_B^{\rm vac}/2\pi\right)^{3/2}$, as explained in \cite{Coleman}.
From this, the false vacuum decay rate (the decay rate per unit time) for a large volume $\Omega$ is
\beq
\label{eq:Gavac}
\Ga^{\rm vac}=\Omega A^{\rm vac} \left(\frac{S_B^{\rm vac}}{2\pi}\right)^{3/2}e^{-S_B^{\rm vac}},
\eeq
where, as in (\ref{eq:Gathin}), $A^{\rm vac}$ comes from the determinant (again excluding zero modes) arising in the saddle-point evaluation of the path integral.

We cannot directly compare \eqref{eq:Gathin1} and \eqref{eq:Gavac}, of course, because we imagine that the volume $\Omega$ contains a large number of vortices (the density of which depends on details of the cosmological phase transition giving rise to them and on the subsequent expansion of the universe until vacuum decay occurs), and also because a universe containing vortices could decay via vortex tunneling or via ordinary vacuum decay with bubble nucleation far from any vortex (assuming the vortices are well-separated). However we can say that the presence of vortices has two effects on vacuum decay: first, it reduces the volume available for ordinary vacuum decay (which presumably must happen sufficiently far from a vortex); secondly, it allows for decay via vortex tunneling. While a detailed analysis is probably fairly involved, given that the vortex bounce action \eqref{eq:thinbounceaction4} is greater than the vacuum bounce action \eqref{eq:SBvac}, it seems likely that vortices impede vacuum decay rather than speeding it up. For instance, if we simply neglect the contribution of ordinary vacuum decay to the decay of a gas of vortices (a reasonable approximation if the density of vortices is high enough) then we can compare the two rates: if $N$ vortices are in the volume $\Omega$, we find
 \[
\frac{\Ga^{\rm vac}}{N\Ga^{\rm thin}} = \frac{ \Omega A^{\rm vac}\left(\frac{S_B^{\rm vac}}{2\pi}\right)^{3/2}\, e^{-\frac{\pi}{12\eps^2}} }
{ N A^{\rm thin}\left(\frac{S_B^{\rm thin}}{2\pi}\right)^{1/2}\, e^{-\frac{4\sqrt{2}\pi}{15\eps^2}} }
= \frac{ \Omega A^{\rm vac} } {N A^{\rm thin} }\, \frac{\sqrt 5 }{2^{1/4}96\eps^2}\, 
e^{\left( \frac{4\sqrt{2}}{15} - \frac{1}{12} \right)\frac{\pi}{\eps^2}}.
\]
Unfortunately, the exponential factors work strongly against the speed-up of vacuum decay by vortices since the last factor is $\exp({\rm (positive)}/\eps^2)$, which (recalling that we have assumed $\eps\ll1$ from the beginning) is exponentially large.

It would be useful if we could estimate the value of $N$ for a volume $\Omega$ or in other words, the density of the vortices
\beq
\rho=\frac{N}{\Omega}.
\eeq
Standard analysis for estimating the density of topological defects during phase transition is based on work of Kibble \cite{kibble} and Zurek \cite{zurek}, however this analysis depends strongly on the model that describes the phase transition.  Our model is reliable only well after the phase transition and therefore should not be used to estimate the vortex density.  If this density is sufficiently large in the parameters of the Lagrangian, then the vortex induced decay rate could in fact  dominate over the vacuum bubble decay rate.

An important observation is that ordinary vacuum decay depends only on the parameters of the potential ($\epsilon$ here), whereas in principle vortex tunneling depends also on the winding number of the vortex and on $e$. (These dependences happened to cancel in \eqref{eq:thinbounceaction4}, but they do not cancel generally.) This suggests that we examine parameters where the vortex tunneling is sped up (by reducing the tunneling barrier between ${\hR}_0$ and ${\hR}_1$ in Fig.~\ref{fig:Ehat}), a situation to which we now turn our attention.

\subsection{Thin-wall vortices and the dissociation point}\label{subsec-dissociation}

It is clear from the discussion following \eqref{eq:Ehat} that within the thin-wall approximation, changing parameters of the model can lead to an effective energy functional for which vortices are no longer classically stable; this occurs for $\hat\eps>{\hat\eps}_c$ as first analyzed in \cite{uj}.  We call the critical point $\hat\eps={\hat\eps}_c$ the dissociation point.  If we approach this point from the side of stable classical vortices, they will simply dissociate and trigger the conversion of the false vacuum to true vacuum without any suppression. For $\hat\eps\ltwid{\hat\eps}_c$, the suppression will be tiny and we expect vortices, if present, to have a dramatic effect on the stability of the vacuum.  

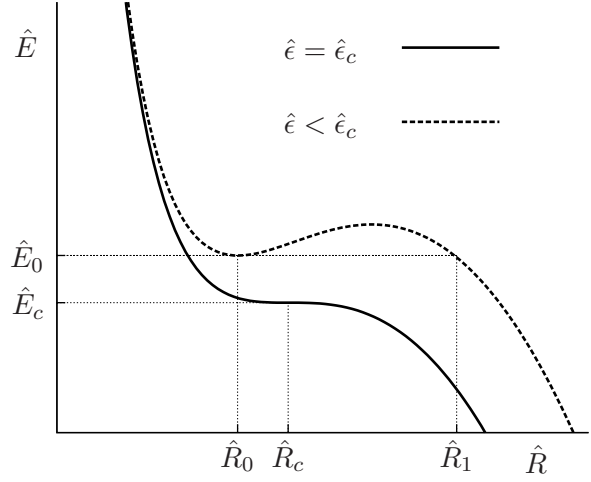
\begin{figure}[ht]
\begin{center}
\input{fig9-rm.tex}
\end{center}
\caption{$\hE(\hR)$ for $\hat\epsilon={\hat\epsilon}_c$ and for $\hat\epsilon<{\hat\epsilon}_c$.
} \label{fig:fig9}
\end{figure}

To study this effect, we must evaluate the action \eqref{eq:thinbounceaction2} to leading order as $\hat\eps\to{{\hat\eps}_c}^-$.  In this limit (see Fig.~\ref{fig:fig9}), $\hR_0(\hat\eps)$ and $\hR_1(\hat\eps)$ approach $\hR_c$ and $\hE_0(\hat\eps)$ approaches $\hE_c$.

Indeed, in this limit we can write
\beq
\hE(\hR)=\frac{1}{2{\hR}^2} + \hR - \hat \eps {\hR}^2
\simeq \hE_0+c(\hR-\hR_0)^2(\hR_1-\hR)
\label{eq:hE}
\eeq
for $\hR$ near $\hR_c$, where $\hR_0$, $\hR_1$, $\hE_0$, and $c$ are functions of $\eps$. Let $\alpha$ parameterize the approach to the dissociation point:
\beq
\label{eq:alpha}
\hat\eps={\hat\eps}_c(1-\alpha).
\eeq
Then we can write
\[
\hR_0=\hR_c(1-\delta_0),\qquad
\hR_1=\hR_c(1+\delta_1),\qquad
\hE_0=\hE_c(1+\delta_2),\qquad
\]
where we expect $\delta_{0,1,2}$ to go to zero as $\alpha\to0$. By expanding the two expressions for $\hE$ given in \eqref{eq:hE} in powers of $\hR-\hR_c$, we can calculate $\delta_{0,1,2}$ as well as $c$. This is somewhat tedious but straightforward; we find
\[
\delta_0=\sqrt{\frac{{\hat\eps}_c\hR_c^4}{3}\alpha},\qquad
\delta_1=2\sqrt{\frac{{\hat\eps}_c\hR_c^4}{3}\alpha},\qquad
\delta_2=\frac{{\hat\eps}_c\hR_c^2}{\hE_c}\alpha,\qquad
c=\frac{2}{\hR_c^5},
\]
or, using the values for ${\hat\eps}_c$, $\hR_c$ and $\hE_c$ given earlier,
\[
\delta_0=\sqrt{\frac{\alpha}{2}},\qquad
\delta_1=\sqrt{2\alpha},\qquad
\delta_2=\frac{\alpha}{2},\qquad
c=\frac{1}{2^{7/3}}.
\]

To evaluate the action \eqref{eq:thinbounceaction1}, we rewrite the last term as
\[
\frac12 - \hE_0{\hR}^2 + {\hR}^3 - \hat\eps{\hR}^4=
\hR^2(\hE(\hR)-\hE_0)=\hR^2 c(\hR-\hR_0)^2(\hR_1-\hR),
\]
giving
\[
S_B^{\rm thin} = 2\pi\left(\frac{2n}{e}\right)^{4/3}
\int_{{\hR}_0}^{{\hR}_1}\frac{d\hR}{{\hR}}
\sqrt{
\left(\frac14+\frac{{\hR}^3}{2}\right)
c(\hR-\hR_0)^2(\hR_1-\hR)
}
\]
Now, for small $\alpha$, the region of integration is small and we can approximate $\hR\simeq\hR_c$ except in the last two factors, giving
\[
S_B^{\rm thin} \simeq2 \pi\left(\frac{2n}{e}\right)^{4/3}
\sqrt{c\hR_c^3}\sqrt{\frac14+\frac{{\hR_c}^3}{2}}
\int_{-\delta_0}^{\delta_1}dr
(r+\delta_0)\sqrt{\delta_1-r},
\]
where we have gone to the integration variable $r$ defined by $\hR=\hR_c(1+r)$.
The integral is standard and is equal to $(4/15)(\delta_0+\delta_1)^{5/2}$; substituting the values given earlier for the parameters we find
\beq
\label{eq:dissociation1}
S_B^{\rm thin} \simeq 2\pi\left(\frac{2n}{e}\right)^{4/3}
\frac{ 2^{-5/12}{3^{5/2}}}{5} \alpha^{5/4}
\eeq
which goes to zero as $\alpha\to0$, which is the dissociation point, as anticipated in \cite{uj} and also discussed in \cite{uk}.

This result is interesting.  If we imagine a supersymmetric theory spontaneously breaking at an intermediate high energy length scale to a broken abelian symmetry, we expect that there will be vortex lines trapped in the universe.  As it cools, the broken symmetry is restored and the universe is prone to vacuum decay.  This decay due to the usual bubbles as analyzed by Coleman \cite{Coleman}  is generically exponentially suppressed.  However, as the universe cools, the coupling constant associated with the broken abelian gauge theory in principle renormalizes in the opposite fashion to an asymptotically free theory.  Therefore the abelian gauge coupling constant $e$ decreases as the universe cools.  As it decreases, $\hat\epsilon=(2n/e)^{2/3}\epsilon$ increases and if $\hat\epsilon\rightarrow{\hat\eps}_c$, the vortex lines will simply dissociate.  Indeed, as the coupling constant decreases, the tunneling amplitude is unsupressed as is evident from \eqref{eq:dissociation1}.  

It is important to underline that the energy of our vortices behaves like $n^{2/3}$.  This implies that the broken vacuum is in fact analogous to a Type I superconductor \cite{ashmer}.  One vortex of large number of flux quanta $n$   is energetically favoured to $n$ vortices each of only one flux quantum.  Clearly the energy of the latter is linearly proportional to $n$, $E_{{ n\,\rm single\, flux\, vortices}}\sim n$ which is always greater than $E_{{\rm single\, vortex \,of\, flux}\, n}\sim n^{2/3}$ that we find for thin walled vortices.  Therefore in our model, if there is trapped magnetic field in the vacuum, then it will be segregated in one or a few  vortices of large total magnetic flux in each.  These vortices will be necessarily thin walled, exactly as is required for our analysis, and will promote decay of the vacuum.  

In fact, to compute the actual decay of vortex lines in a three dimensional context (as opposed to the vortices considered in this work) requires a more detailed analysis, involving an effective field corresponding to the radius of the thin-walled cosmic string as a function of the Euclidean time and the spatial coordinate along the string, rather than the simple 2+1 dimensional analysis presented here.  That analysis will appear in a forthcoming paper \cite{llmpyy}.

\section{Discussion}\label{sec-discussion}

We have discussed the possibility that vortices in a universe trapped in a symmetry-breaking false vacuum can have a significant effect on vacuum decay. If a classically stable vortex decays into a large vortex which is classically unstable, this vortex will expand, leaving behind it the symmetry-preserving vacuum. It is possible that conventional vacuum decay proceeds by tunneling and is exponentially suppressed, while vortex tunneling is unsuppressed or is only slightly suppressed. In this case, the presence of vortices in the universe would catalyze vacuum decay.  As we have seen, this occurs as one approaches the dissociation limit \cite{pjs}, in which the tunneling barrier between a classically stable vortex and an unstable, expanding vortex shrinks away (see Fig.~\ref{fig:fig9}).  Thus there is a range of parameter space for which the vortices are classically stable, but trigger the decay of the false vacuum in an essentially unsuppressed manner.  

If we imagine a symmetry breaking scenario where a hot initial state condenses and is trapped in the false vacuum state, evidently over causally disconnected regions, the order parameter can be pointing in different directions.  As the state cools, this gives rise to an intermediate state of essentially the false vacuum, punctuated by quantum mechanically unstable vortices.  If the decay of these vortices is essentially unsuppressed, the false vacuum is quickly converted to true vacuum via the tunneling that we have described and the liberation of the trapped flux in the vortices.  

Our analysis could find applications in condensed matter situations, for example in the transition from the $A$  to $B$ phase of superfluid $^3$He where it is observed that a phase transition occurs many orders of magnitude faster than the expected quantum mechanical decay rate, currently an open problem \cite{legg}. Secondly, a type-II superconductor in the intermediate-field region (between the upper and lower critical fields \cite{ashmer}) is penetrated by vortices. A superconductor described by the model discussed in this paper (or one with similar features) would be unstable, but could be extremely long-lived. However, the presence of vortices, if the model is near its dissociation point, would destabilize the superconductivity.

\section*{Acknowledgements}
This work was financially supported in part by the National Research Foundation of Korea grant funded by the Ministry of Education, Science and Technology through the Center for Quantum Spacetime (CQUeST) of Sogang University (2005-0049409), by the Natural Science and Engineering Council of Canada,  by a Department of Science and Technology Grant, India, by  the Coopération Québec-Maharashtra (Inde) program of the Ministère des relations internationales du Québec and by the Direction de relations internationales de l'Université de Montréal. WL was supported by the Basic Science Research Program through the National Research Foundation of Korea (NRF) funded by the Ministry of Education, Science and Technology (2012R1A1A2043908).  DY is supported by the JSPS Grant-in-Aid for Scientiﬁc Research (A) No. 21244033. RM thanks McGill University and MP thanks the Perimeter Institute for Theoretical Physics for hospitality while this work was in progress.  We thank Marie-Lou Gendron-Marsolais and Yan Gobeil for verifying much of the numerical work.

\end{document}

%% file: Ehat-take2.tex
% GNUPLOT: LaTeX picture with Postscript
\begingroup
  \makeatletter
  \providecommand\color[2][]{%
    \GenericError{(gnuplot) \space\space\space\@spaces}{%
      Package color not loaded in conjunction with
      terminal option `colourtext'%
    }{See the gnuplot documentation for explanation.%
    }{Either use 'blacktext' in gnuplot or load the package
      color.sty in LaTeX.}%
    \renewcommand\color[2][]{}%
  }%
  \providecommand\includegraphics[2][]{%
    \GenericError{(gnuplot) \space\space\space\@spaces}{%
      Package graphicx or graphics not loaded%
    }{See the gnuplot documentation for explanation.%
    }{The gnuplot epslatex terminal needs graphicx.sty or graphics.sty.}%
    \renewcommand\includegraphics[2][]{}%
  }%
  \providecommand\rotatebox[2]{#2}%
  \@ifundefined{ifGPcolor}{%
    \newif\ifGPcolor
    \GPcolorfalse
  }{}%
  \@ifundefined{ifGPblacktext}{%
    \newif\ifGPblacktext
    \GPblacktexttrue
  }{}%
  % define a \g@addto@macro without @ in the name:
  \let\gplgaddtomacro\g@addto@macro
  % define empty templates for all commands taking text:
  \gdef\gplbacktext{}%
  \gdef\gplfronttext{}%
  \makeatother
  \ifGPblacktext
    % no textcolor at all
    \def\colorrgb#1{}%
    \def\colorgray#1{}%
  \else
    % gray or color?
    \ifGPcolor
      \def\colorrgb#1{\color[rgb]{#1}}%
      \def\colorgray#1{\color[gray]{#1}}%
      \expandafter\def\csname LTw\endcsname{\color{white}}%
      \expandafter\def\csname LTb\endcsname{\color{black}}%
      \expandafter\def\csname LTa\endcsname{\color{black}}%
      \expandafter\def\csname LT0\endcsname{\color[rgb]{1,0,0}}%
      \expandafter\def\csname LT1\endcsname{\color[rgb]{0,1,0}}%
      \expandafter\def\csname LT2\endcsname{\color[rgb]{0,0,1}}%
      \expandafter\def\csname LT3\endcsname{\color[rgb]{1,0,1}}%
      \expandafter\def\csname LT4\endcsname{\color[rgb]{0,1,1}}%
      \expandafter\def\csname LT5\endcsname{\color[rgb]{1,1,0}}%
      \expandafter\def\csname LT6\endcsname{\color[rgb]{0,0,0}}%
      \expandafter\def\csname LT7\endcsname{\color[rgb]{1,0.3,0}}%
      \expandafter\def\csname LT8\endcsname{\color[rgb]{0.5,0.5,0.5}}%
    \else
      % gray
      \def\colorrgb#1{\color{black}}%
      \def\colorgray#1{\color[gray]{#1}}%
      \expandafter\def\csname LTw\endcsname{\color{white}}%
      \expandafter\def\csname LTb\endcsname{\color{black}}%
      \expandafter\def\csname LTa\endcsname{\color{black}}%
      \expandafter\def\csname LT0\endcsname{\color{black}}%
      \expandafter\def\csname LT1\endcsname{\color{black}}%
      \expandafter\def\csname LT2\endcsname{\color{black}}%
      \expandafter\def\csname LT3\endcsname{\color{black}}%
      \expandafter\def\csname LT4\endcsname{\color{black}}%
      \expandafter\def\csname LT5\endcsname{\color{black}}%
      \expandafter\def\csname LT6\endcsname{\color{black}}%
      \expandafter\def\csname LT7\endcsname{\color{black}}%
      \expandafter\def\csname LT8\endcsname{\color{black}}%
    \fi
  \fi
  \setlength{\unitlength}{0.0500bp}%
  \begin{picture}(7200.00,5040.00)%
    \gplgaddtomacro\gplbacktext{%
      \csname LTb\endcsname%
      \put(2088,2018){\makebox(0,0)[r]{\strut{}${\hat E}_0$}}%
      \put(2622,886){\makebox(0,0){\strut{}${\hat R}_0$}}%
      \put(5232,886){\makebox(0,0){\strut{}${\hat R}_1$}}%
      \put(1664,3717){\makebox(0,0)[l]{\strut{}$\hat E(\hat R)$}}%
      \put(5990,893){\makebox(0,0)[l]{\strut{}$\hat R$}}%
    }%
    \gplgaddtomacro\gplfronttext{%
    }%
    \gplbacktext
    \put(0,0){\includegraphics{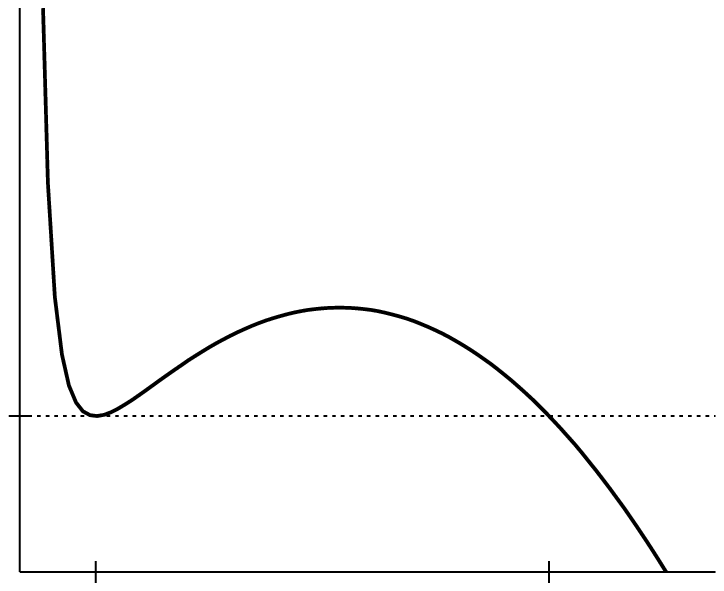}}%
    \gplfronttext
  \end{picture}%
\endgroup

%% file: fig9-rm.tex
% GNUPLOT: LaTeX picture with Postscript
\begingroup
  \makeatletter
  \providecommand\color[2][]{%
    \GenericError{(gnuplot) \space\space\space\@spaces}{%
      Package color not loaded in conjunction with
      terminal option `colourtext'%
    }{See the gnuplot documentation for explanation.%
    }{Either use 'blacktext' in gnuplot or load the package
      color.sty in LaTeX.}%
    \renewcommand\color[2][]{}%
  }%
  \providecommand\includegraphics[2][]{%
    \GenericError{(gnuplot) \space\space\space\@spaces}{%
      Package graphicx or graphics not loaded%
    }{See the gnuplot documentation for explanation.%
    }{The gnuplot epslatex terminal needs graphicx.sty or graphics.sty.}%
    \renewcommand\includegraphics[2][]{}%
  }%
  \providecommand\rotatebox[2]{#2}%
  \@ifundefined{ifGPcolor}{%
    \newif\ifGPcolor
    \GPcolorfalse
  }{}%
  \@ifundefined{ifGPblacktext}{%
    \newif\ifGPblacktext
    \GPblacktexttrue
  }{}%
  % define a \g@addto@macro without @ in the name:
  \let\gplgaddtomacro\g@addto@macro
  % define empty templates for all commands taking text:
  \gdef\gplbacktext{}%
  \gdef\gplfronttext{}%
  \makeatother
  \ifGPblacktext
    % no textcolor at all
    \def\colorrgb#1{}%
    \def\colorgray#1{}%
  \else
    % gray or color?
    \ifGPcolor
      \def\colorrgb#1{\color[rgb]{#1}}%
      \def\colorgray#1{\color[gray]{#1}}%
      \expandafter\def\csname LTw\endcsname{\color{white}}%
      \expandafter\def\csname LTb\endcsname{\color{black}}%
      \expandafter\def\csname LTa\endcsname{\color{black}}%
      \expandafter\def\csname LT0\endcsname{\color[rgb]{1,0,0}}%
      \expandafter\def\csname LT1\endcsname{\color[rgb]{0,1,0}}%
      \expandafter\def\csname LT2\endcsname{\color[rgb]{0,0,1}}%
      \expandafter\def\csname LT3\endcsname{\color[rgb]{1,0,1}}%
      \expandafter\def\csname LT4\endcsname{\color[rgb]{0,1,1}}%
      \expandafter\def\csname LT5\endcsname{\color[rgb]{1,1,0}}%
      \expandafter\def\csname LT6\endcsname{\color[rgb]{0,0,0}}%
      \expandafter\def\csname LT7\endcsname{\color[rgb]{1,0.3,0}}%
      \expandafter\def\csname LT8\endcsname{\color[rgb]{0.5,0.5,0.5}}%
    \else
      % gray
      \def\colorrgb#1{\color{black}}%
      \def\colorgray#1{\color[gray]{#1}}%
      \expandafter\def\csname LTw\endcsname{\color{white}}%
      \expandafter\def\csname LTb\endcsname{\color{black}}%
      \expandafter\def\csname LTa\endcsname{\color{black}}%
      \expandafter\def\csname LT0\endcsname{\color{black}}%
      \expandafter\def\csname LT1\endcsname{\color{black}}%
      \expandafter\def\csname LT2\endcsname{\color{black}}%
      \expandafter\def\csname LT3\endcsname{\color{black}}%
      \expandafter\def\csname LT4\endcsname{\color{black}}%
      \expandafter\def\csname LT5\endcsname{\color{black}}%
      \expandafter\def\csname LT6\endcsname{\color{black}}%
      \expandafter\def\csname LT7\endcsname{\color{black}}%
      \expandafter\def\csname LT8\endcsname{\color{black}}%
    \fi
  \fi
  \setlength{\unitlength}{0.0500bp}%
  \begin{picture}(7200.00,5040.00)%
    \gplgaddtomacro\gplbacktext{%
      \csname LTb\endcsname%
      \put(2088,2100){\makebox(0,0)[r]{\strut{}${\hat E}_c$}}%
      \put(2088,2456){\makebox(0,0)[r]{\strut{}${\hat E}_0$}}%
      \put(3546,949){\makebox(0,0){\strut{}${\hat R}_0$}}%
      \put(3927,949){\makebox(0,0){\strut{}${\hat R}_c$}}%
      \put(5197,949){\makebox(0,0){\strut{}${\hat R}_1$}}%
      \put(1863,4042){\makebox(0,0)[l]{\strut{}$\hat E$}}%
      \put(5710,904){\makebox(0,0)[l]{\strut{}$\hat R$}}%
    }%
    \gplgaddtomacro\gplfronttext{%
      \csname LTb\endcsname%
      \put(4448,4024){\makebox(0,0)[r]{\strut{}$\hat\epsilon={\hat\epsilon}_c$}}%
      \csname LTb\endcsname%
      \put(4448,3464){\makebox(0,0)[r]{\strut{}$\hat\epsilon<{\hat\epsilon}_c$}}%
    }%
    \gplbacktext
    \put(0,0){\includegraphics{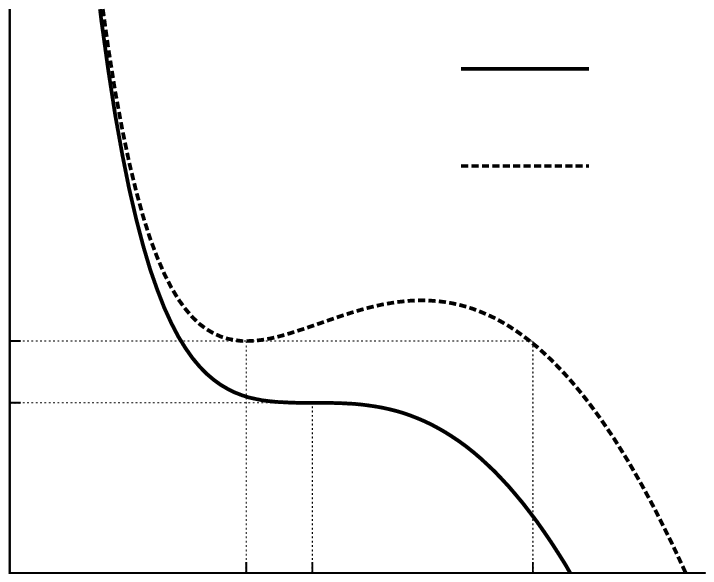}}%
    \gplfronttext
  \end{picture}%
\endgroup